\documentclass[12pt]{iopart}
\usepackage{iopams}
\usepackage{epsf}
\begin{document}

\title[Self-force Regularization]{Self-force Regularization 
in the Schwarzschild Spacetime}

\author{Wataru Hikida,\dag\ Hiroyuki Nakano\ddag\ and Misao Sasaki\dag  
}

\address{\dag\ Yukawa Institute for Theoretical Physics, 
Kyoto University, Kyoto, Japan}

\address{\ddag\ Department of Mathematics and Physics, 
Osaka City University, Osaka, Japan}

\eads{\mailto{hikida@yukawa.kyoto-u.ac.jp}, 
\mailto{denden@sci.osaka-cu.ac.jp}, 
\mailto{misao@yukawa.kyoto-u.ac.jp}}

\begin{abstract}

We discuss the gravitational self-force on a particle in a black hole 
space-time. For a point particle, the full (bare) self-force diverges. 
The metric perturbation induced by a particle can be divided into two parts, 
the direct part (or the S part) and the tail part (or the R part), 
in the harmonic gauge, and the regularized self-force is derived from
 the R part which is regular and satisfies the source-free perturbed
 Einstein equations. 
But this formulation is abstract, so when we apply to black hole-particle
 systems, 
there are many problems to be overcome in order to derive a 
concrete self-force. 
These problems are roughly divided into two parts. 
They are the problem of regularizing the divergent self-force, 
i.e., ``subtraction problem'' and the problem of the singularity 
in gauge transformation, i.e., ``gauge problem''. 
In this paper, we discuss these problems in the Schwarzschild background
and report some recent progress. 
\end{abstract}

\pacs{04.70.Bw, 04.25.Nx\hfill \today}

\submitto{\CQG}

\maketitle

\section{Introduction}

Thanks to recent advances in technology, an era of gravitational 
wave astronomy has almost arrived. There are already several large-scale 
laser interferometric gravitational wave detectors that are in operation 
in the world. Among them are TAMA300 \cite{TAMA}, LIGO \cite{LIGO}, 
GEO-600 \cite{GEO}, and VIRGO \cite{VIRGO} is expected to start 
its operation soon. The primary targets of these ground-based detectors 
are inspiralling compact binaries, which are expected to be detected 
in the near future.

There are also space-based interferometric detector projects.
LISA is now on its R \& D stage~\cite{LISA},
and there is a future plan called DECIGO/BBO~\cite{DECIGO,BBO}.
These space-based detectors can detect gravitational waves
from solar-mass compact objects orbiting supermassive black holes.
To extract out physical information of such binary systems,
it is essential to know the theoretical gravitational waveforms
with sufficient accuracy.
The black hole perturbation approach is most suited for
this purpose. In this approach, one considers gravitational waves
emitted by a point particle that represents a compact
object orbiting a black hole, assuming the mass of the particle ($\mu$)
is much less than that of the black hole ($M$); $\mu\ll M$.

In the lowest order in the mass ratio $(\mu/M)^0$,
the orbit of the particle is a geodesic on the background geometry
of a black hole. Already in this lowest order,
by combining with the assumption of adiabatic orbital evolution,
this approach has been proved to be very powerful for
evaluating general relativistic corrections to the gravitational
waveforms, even for neutron star-neutron star
(NS-NS) binaries where the assumption of this approach is maximally
violated~\cite{MSSTT}.

In the next order, the orbit deviates from the geodesic
because the spacetime is perturbed by
the particle. We can interpret this deviation as the effect 
of the self-force of the particle on itself. Since it is essential 
to take account of this deviation to predict the orbital evolution
accurately, we have to derive the equation of motion that 
includes the self-force. 
The gravitational self-force is, however, not easily obtainable.
There are two main obstacles. 

First, the full (bare) metric perturbation 
diverges at the location of the particle, which is point-like,
hence so does the self-force. The full metric perturbation can be formally
divided into two parts; the direct part that comes from the perturbation
propagating along the light cone of the background spacetime directly from
the source particle, and the tail part that arises from the curvature
scattering of the perturbation, which exists within the light cone.
The self-force is given by the tail part
of the metric perturbation which is regular at the location of the
 particle~\cite{reaction,QuiWal}.
Recently, this is reformulated in a more elaborated way 
by Detweiler and Whiting~\cite{Detweiler:2002mi}. In this new formulation,
the direct part is replaced by the S part and the tail part is replaced by 
the R part. The R part is equivalent to the tail
part as far as the self-force calculation is concerned, but which has a much
better property that it is a solution of source-free linearized Einstein
equations. Thus, one has to identify 
the R part of the metric perturbation to obtain a meaningful self-force. 
However, by construction, the R part cannot be determined locally but depends
on the whole history of the particle. Therefore, one usually identifies 
the direct part (or the S part) which can be evaluated locally
in the vicinity of the particle to a necessary order, i.e.,
by local analysis, and subtract it from the full metric perturbation.
This identification of the S part is {\it the subtraction problem}.

Second, the regularized self-force is formally defined 
only in the harmonic (Lorentz) gauge,
in which linearized Einstein equations are hyperbolic, which 
enables us to define the S part and the R part uniquely.
On the other hand, the metric perturbation of a 
black hole geometry can be calculated only in the radiation gauge
in the Kerr background~\cite{Chrzanowski:wv}, or in the 
Regge-Wheeler(-Zerilli) gauge in the Schwarzschild
 background~\cite{RegWhe,Zer}. Hence, one has to find a 
gauge transformation to express the full metric perturbation 
and the S part in a same gauge, to identify the R part
in that gauge correctly. This is {\it the gauge problem}. 

As a method to solve the subtraction problem,
extending earlier work~\cite{Mino,MinNak}, Mino, Nakano and Sasaki
formulated two types of regularization methods,
namely, the `power expansion regularization' and 
the `mode-decomposition regularization'~\cite{Nakano:2001kw,Mino:2001mq}. 
In particular, the mode-decomposition regularization was found
to be quite powerful. Meanwhile, Barack and Ori independently developed
the `mode-sum regularization'~\cite{Mode-sum,Barack:2002mh,Barack:2002bt}.
Although the mode-decomposition regularization and the mode-sum 
regularization were quite different in their formulations,
they are essentially the same in the sense that the spherical harmonic
expansion is used to regularize the divergent metric.
In fact, it was shown that the two methods give the same result
for the S part in the harmonic gauge, in the form of a (divergent) 
spherical harmonic series, for an arbitrary orbit in the Schwarzschild 
background~\cite{Barack:2001gx}. 

However, it seems extremely difficult to solve 
the metric perturbation under the harmonic gauge 
 because the metric components couple to each other in 
a complicated way~\cite{Sago:2002fe}. This is one of the reasons 
why the gauge problem is difficult to solve. 
Recently, Barack and Ori~\cite{Barack:2001ph} gave a useful 
insight into the gauge problem. They proposed an intermediate gauge 
approach in which only the direct part of the metric 
in the harmonic gauge is subtracted from the full metric 
perturbation in the Regge-Wheeler (RW) gauge.
They then argued that the gauge-dependence of the self-force 
is unimportant when averaged over a sufficiently long lapse of time. 
Using this approach, the gravitational self-force for an orbit 
plunging into a Schwarzschild black hole was calculated 
by Barack and Lousto~\cite{Barack:2002ku}. 

In the case of the Kerr background,
which is of our ultimate interest, there is additionally 
a much more intricate 
technical issue of how to treat the spheroidal harmonics:
Neither the eigenfunction nor the eigenvalue has 
simple analytical expressions, and they are entangled with 
the frequency eigenvalues of Fourier modes. 
Further, the only known gauge in 
which the metric perturbation can be evaluated is the radiation 
gauge formulated by Chrzanowski~\cite{Chrzanowski:wv}.
However, the Chrzanowski construction of the metric perturbation 
becomes ill-defined in the neighborhood of the particle, 
i.e., the Einstein equations are not satisfied there~\cite{Barack:2001ph}. 
Some progress was made by Ori~\cite{Ori:2002uv} to obtain the correct,
full metric perturbation in the Kerr background. The regularization 
parameters in the mode-sum regularization for the Kerr case are 
calculated by Barack and Ori~\cite{Barack:2002mh2}. 
Although our ultimate goal is to overcome these difficulties altogether, 
since each one is sufficiently involved, we choose to
proceed step by step, and focus on the Schwarzschild background. 

In this paper, in order to obtain the gravitational self-force, 
we consider a particle orbiting a Schwarzschild black hole, 
and propose a method to calculate the regularized self-force by 
solving the subtraction and gauge problems simultaneously. 
Namely, we develop a method to regularize the self-force 
in the Regge-Wheeler gauge. 
The regularization is done by the mode-decomposition regularization, 
and we introduce the concept of a ``finite gauge transformation''
of the S part~\cite{Minopc,Nakano:2003he}.
It is noted that, although our approach is philosophically quite
different from the intermediate gauge approach by
Barack and Or~\cite{Barack:2001ph},
practically both approaches turn out to give the same result 
as far as the calculation of the gauge transformation
is concerned. 

The paper is organized as follows. In Section~\ref{sec:MDR}, 
we review the mode-decomposition regularization.
We first derive the direct part 
of the self-force (which is equivalent to the S part
in the coincidence limit of a field point with a point on the
 particle trajectory) by
 local analysis. Then we transform it to an infinite series
of spherical harmonic modes and derive the regularization
counter terms for a general orbit. 
In Section~\ref{sec:gauge}, further focusing on a circular
orbit, we perform a gauge transformation from the harmonic
gauge to the RW gauge, 
and present the first post-Newtonian order self-force.
One of the reasons why we had to focus on circular orbits
is that it is formidable to obtain a closed analytic expression
of the full metric perturbation for a general orbit even
under the post-Newtonian expansion.
In Section~\ref{seq:newreg}, turning back to a simpler case
of the self-force of a scalar-charged particle,
we present a new regularization method that can apply to
an arbitrary orbit. This new method is a
variation (or an improved version) of the mode-decomposition
regularization. The essential point is to divide the
full field in the frequency domain into two parts,
the $\tilde{\rm S}$ part and $\tilde{\rm R}$ part,
in such a way that the divergent part is totally contained
in the $\tilde{\rm S}$ part whose frequency integral can be
 analytically performed.

\section{Mode-decomposition regularization}\label{sec:MDR}

In this section, we review the mode-decomposition regularization
 method, and derive the regularization counter terms for an
general orbit in the Schwarzschild background.
For simplicity, we consider a scalar-charged particle
which sources a scalar field $\varphi$. The extension to
the gravitational case is straightforward. The only complication
is the presence of additional tensor indices associated with the
 metric perturbation.

Since the tail part depends non-locally on the geometry of a
background spacetime, it is almost impossible to calculate it
directly. However, for a certain class of spacetimes such as 
Schwarzschild/Kerr geometries, there is a way to calculate 
the full field generated by a point source.
Considering a field point slightly off the particle trajectory, 
it is then possible to obtain the tail part by subtracting 
the locally given divergent direct part from the full field. 

Let $x=z(\tau)$ be a trajectory of the particle with $\tau$ 
beging the proper time along the trajectory.
Then the field $\varphi(x)$ generated by the particle is
given by
\begin{eqnarray}
\varphi^{\rm full}(x) &=& -q\int d\tau\, G^{\rm full}(x,z(\tau)) \,,
\label{eq:full-g}
\end{eqnarray}
where $G^{\rm full}(x,x')$ is the retarded Green function
satisfying the Klein-Gordon equation,
\begin{eqnarray}
\nabla^\alpha \nabla_\alpha G^{\rm full}(x,x') &=&
-{\delta^{(4)}(x-x')\over \sqrt{-g}} \,. 
\label{eq:wave}
\end{eqnarray}
The self-force at $x=z(\tau_0)$ is schematically given by 
\begin{eqnarray}
F_\alpha(\tau_0) &=&
\lim_{x\rightarrow z(\tau_0)}F_\alpha[{\varphi}^{\rm tail}](x)
\,, \label{eq:scheme}
\end{eqnarray}
where
\begin{eqnarray}
F_\alpha[{\varphi}^{\rm tail}](x) &=&
F_\alpha[{\varphi}^{\rm full}](x)-F_\alpha[{\varphi}^{\rm dir}](x)
\,, \quad (x \not = z(\tau)) \,.
\nonumber
\end{eqnarray}
The symbols ${\varphi}^{\rm full}$, ${\varphi}^{\rm dir}$ and
${\varphi}^{\rm tail}$ stand for the full field, the direct part 
and the tail part, respectively.~\footnote{In this section,
 we consider the direct part
instead of the S part, because the direct part has a slightly
simpler form for local analysis.}
Both ${\varphi}^{\rm full}$ and ${\varphi}^{\rm dir}$ diverge 
in the coincidence limit $x\to z(\tau)$, while ${\varphi}^{\rm tail}$
is finite in the coincidence limit.
$F_\alpha[...]$ is a tensor operator on the field, and is defined as
\begin{eqnarray}
F_\alpha[{\varphi}] &=& 
q P_\alpha{}^\beta \nabla_\beta \varphi\,,
\label{eq:force}
\end{eqnarray}
where $P_\alpha{}^\beta = \delta_\alpha{}^\beta +V_\alpha V^\beta$
is the projection tensor with
$V^\alpha$ being an appropriate extension of
the four velocity $v^\alpha(\tau_0)$ off the orbital point.

In what follows, we use the Boyer-Lindquist coordinates
for the Schwarzschild background,
\begin{eqnarray}
ds^2 &=&
-\left(1-{2M\over r}\right)dt^2+\left(1-{2M\over r}\right)^{-1} dr^2
+r^2\left(d\theta^2+\sin^2\theta d\phi^2\right) \,,
\end{eqnarray}
and denote a field point by $x=\{t,\,r,\,\theta,\,\phi\}$
and a point on the orbit by $z(\tau_0)=z_0=\{t_0,\,r_0,\,\theta_0,\,\phi_0\}$.

\subsection{Basis} \label{subsec:scheme}

Even in the Newtonian limit, the integration of an orbit involves 
an elliptic function. Thus, numerical computations will be necessary 
at some stage of deriving the self-force. However, it will not be easy
to perform the regularization of the divergence numerically.
The idea to overcome this difficulty is to replace the divergence 
by an infinite series, each term of which is finite and analytically
calculable. 

The derivation of the direct part ${\varphi}^{\rm dir}(x)$ is 
one of the main issues in the regularization calculation. 
The direct part of the scalar field is obtained 
by integrating the direct part of the retarded Green function 
with the scalar charge. The direct part of the retarded Green 
function $G^{\rm dir}$ is given in a covariant manner as~\cite{DeWittBrehme}
\begin{eqnarray}
G^{\rm dir}(x,x') &=& -{1\over 4\pi}\, \theta[\Sigma(x),x']
\sqrt{\Delta(x,x')}\delta\bigl(\sigma(x,x')\bigr)
\,, \label{eq:direct-g}
\end{eqnarray}
where $\sigma(x,x')$ is the bi-scalar of half the squared geodesic distance,
$\Delta(x,x')$ is the generalized van Vleck-Morette determinant,
$\Sigma(x)$ is an arbitrary spacelike hypersurface containing $x$,
and $\theta[\Sigma(x),x']=1-\theta[x',\Sigma(x)]$ is equal to $1$
when $x'$ lies in the past of $\Sigma(x)$
and vanishes when $x'$ lies in the future. The physical meaning of 
the direct part is understood by the factor 
$\theta[\Sigma(x),x']\delta\bigl(\sigma(x,x')\bigr)$ 
in Eq.~(\ref{eq:direct-g}). 
Since $\sigma(x,x')$ describes the geodesic distance between $x$ and $x'$, 
the direct part of the Green function becomes non-zero 
only when $x'$ lies on the past lightcone of $x$. 
Hence the direct part describes the effect 
of the waves propagated directly from $x'$ to $x$ 
without scattered by the background curvature. 

For the actual evaluation of the direct part, a couple of methods have 
been proposed.  In Refs.~\cite{Burko,Barack}, the direct part of the field 
is calculated by picking up a limiting contribution in the full Green 
function from the light cone as
\begin{eqnarray}
\varphi^{\rm dir}(x)= \lim_{\epsilon\to+0}
\int^\infty_{\tau_{\rm ret}(x)-\epsilon}d\tau \,
G^{\rm full}(x,z(\tau))S(\tau)
\,,
\end{eqnarray}
where $G^{\rm full}$ is the retarded Green function,
$S(\tau)$ is the scalar charge density,
and $\tau_{\rm ret}(x)$ is the retarded time defined
by the past light cone condition of the field point $x$ as
\begin{eqnarray}
\theta[\Sigma(x),z(\tau_{\rm ret})]
\delta\bigl(\sigma(x,z(\tau_{\rm ret}))\bigr)=0 \,.
\end{eqnarray}
However, the calculation seems rather cumbersome 
when we apply this method to a general orbit.

In Ref.~\cite{MinNak}, the direct part was evaluated using the local 
bi-tensor expansion technique. Using the bi-tensor, 
the direct part is expanded around the particle location as
\begin{eqnarray}
\varphi^{\rm dir}(x) &=&
q\left[{1 \over \sigma_{;\alpha}(x,z(\tau_{\rm ret}))
v^\alpha(\tau_{\rm ret})}\right] +O(y^2) \,,
\label{eq:direct}
\end{eqnarray}
where the letters $\mu$, $\nu$, $\cdots$ are used for the indices
of the field point $x$,
$\alpha$, $\beta$, $\cdots$ for the indices of the orbital point $z$,
and $v^\alpha(\tau)$ is the orbital four velocity at $z(\tau)$.
The order of the local expansion is represented by the
powers of $y$, where $y$ is the coordinate difference between
the field point $x$ and the orbital point $z_0$; $y^\mu=x^\mu-z^\mu_0$.
Because the full force is quadratically divergent,
we must carry out the local bi-tensor expansion of the full field
up through $O(y)$. By evaluating the local coordinate values of 
the relevant bi-tensors, we obtain the local expansion of the full force 
in a given coordinate system. As described in Ref.~\cite{MinNak}, 
this may be done in a systematic manner, and it is possible to obtain 
the explicit form of the divergence for a general orbit. 
However, the problem is how to decompose it into an appropriate 
infinite series. 

\subsection{Mode-decomposition regularization} \label{subsec:mode}

The mode-decomposition regularization method is based
 on the spherical harmonic series expansion~\cite{Mino:2001mq}. 
There is a delicate problem in this approach. 
The exact decomposition calculation usually needs 
the global analytic structure of the field 
so that we can uniquely define each term in the infinite series.
On the other hand, the divergent direct part to be subtracted is
given only by local analysis of the
 field~\cite{reaction,QuiWal,DeWittBrehme,Quinn}. 
Thus the direct part is defined (and known) only in a local neighborhood of
the particle. Because of this, to obtain a harmonic series expression
of the direct part, we need a global extension of the direct part.
But this extension cannot be done uniquely in practice.
Nevertheless, the final result of the self-force should be unique.
Although we have no explicit proof for the uniqueness of the 
regularization counter terms for the self-force,
the uniqueness is intuitively apparent because the extension is
performed in such a way that the local behavior is retained
with sufficiently accuracy.
In addition, it is reassuring to find that the resulting counter terms, 
presented at the end of this section, agree completely
with those obtained in the mode-sum regularization by
Barack and Ori~\cite{Mode-sum,Barack:2002mh,Barack:2002bt,Barack:2001gx}.

The harmonic decomposition is defined on a two-sphere.
 However both the direct field 
and the full field have divergence on the sphere containing
 the particle location,
the mode decomposition is ill-defined on that sphere. 
Therefore, we perform the harmonic decomposition of the direct and full fields 
on a sphere which does not include, but sufficiently close to the orbit. 
The steps in the mode-decomposition regularization are as follows.

\begin{list}{}{}
\item[1)]
We evaluate both the full field and the direct field at
\begin{eqnarray}
x &=& \{t,r,\theta,\phi\} \,,
\end{eqnarray}
where we do not take the coincidence limit of either $t$ or $r$

\item[2)]
We decompose the full force and direct force
into infinite harmonic series as
\begin{eqnarray}
F_\alpha[{\varphi}^{\rm full}](x) &=&
\sum_{\ell m}F_\alpha^{\ell m}[{\varphi}^{\rm full}](x) \,, \\
F_\alpha[{\varphi}^{\rm dir}](x) &=&
\sum_{\ell m}F_\alpha^{\ell m}[{\varphi}^{\rm dir}](x) \,,
\end{eqnarray}
where $F_\alpha[{\varphi}^{\rm full/dir}](x)$ are expanded 
in terms of the spherical harmonics $Y_{\ell m}(\theta,\phi)$ 
with the coefficients dependent on $t$ and $r$. 
For the direct part, the harmonic expansion is done by extending 
the locally defined direct force over to the whole two-sphere
in a way that correctly reproduces the divergent behavior around
the orbital point $z_0$ up to the finite term.

\item[3)]
We subtract the direct part from the full part in each $\ell$, $m$ mode
to obtain
\begin{eqnarray}
F_\alpha^{\ell m}[{\varphi}^{\rm tail}] &=&
(F_\alpha^{\ell m}[{\varphi}^{\rm full}]
-F_\alpha^{\ell m}[{\varphi}^{\rm dir}]) \,.
\end{eqnarray}
Then, we take the coincidence limit $x\to z_0$. 
Here we note that one can exchange the order of the procedure, 
i.e., first take the coincidence limit and then subtract, provided the mode 
coefficients are finite in the coincidence limit.

\item[4)]
Finally, by taking the sum over the modes, we obtain
the self-force,
\begin{eqnarray}
F_\alpha(\tau_0) &=&
\sum_{\ell m}F_\alpha^{\ell m}[{\varphi}^{\rm tail}](z_0) \,.
\end{eqnarray}
\end{list}

\subsection{Decomposition of the direct part} \label{subsec:decomposition}

The advantage of using the expression~(\ref{eq:direct}) for the
direct part is that we have a systematic method for evaluating it.
Here we describe a method to evaluate the bi-tensors 
in a general regular coordinate system. 

Before we consider the local expansion in a given coordinate system,
we calculate the derivative of Eq.~(\ref{eq:direct}),
and derive the direct part of the force with the local bi-tensor expansion
using the equal-time condition, 
\begin{eqnarray}
0 &=& \left[{d\over d\tau}\sigma(x,z(\tau))\right]_{\tau=\tau_{\rm eq}(x)}
\,. \label{eq:equal}
\end{eqnarray}
We define an extension of the four-velocity off the orbit by
\begin{eqnarray}
V^\alpha(x) :=
{\bar g}_{\alpha{\bar \alpha}}(x,z_{\rm eq})v^{\bar\alpha}_{\rm eq}
\,; \quad z_{\rm eq}=z(\tau_{\rm eq}(x)) \,, 
\quad v^{\bar\alpha}_{\rm eq}
=\frac{dz^{\bar\alpha}}{d\tau|_{\tau=\tau_{\rm eq}(x)}} \,,
\label{eq:ppexV}
\end{eqnarray}
where ${\bar g}_{\alpha{\bar \alpha}}$ is the parallel displacement bi-vector.

Using the formulas in Refs.~\cite{reaction,DeWittBrehme}, we have 
\begin{eqnarray}
F_\alpha[\varphi^{\rm dir}](x) &=&
q {\bar g}_\alpha{}^{\bar \alpha}(x,z_{\rm eq}){1\over \epsilon^3\kappa}
\Biggl\{\sigma_{;\bar\alpha}(x,z_{\rm eq})
\nonumber \\ && 
+{1 \over 3}\epsilon^2R_{\bar\alpha\bar\beta\bar\gamma\bar\delta}(z_{\rm eq})
v^{\bar\beta}_{\rm eq}\sigma^{;\bar\gamma}(x,z_{\rm eq})
v^{\bar\delta}_{\rm eq}\Biggr\}+O(y)
\,, \label{eq:dir_force} \\
\epsilon &=& \sqrt{2\sigma(x,z_{\rm eq})}
\,, \\
\kappa &=& \sqrt{-\sigma_{\bar\alpha\bar\beta}(x,z_{\rm eq})
v^{\bar\alpha}_{\rm eq}v^{\bar\beta}_{\rm eq}}
\nonumber \\
&=& 1+{1\over 6}R_{\bar\alpha\bar\beta\bar\gamma\bar\delta}(z_{\rm eq})
v^{\bar\alpha}_{\rm eq}\sigma^{;\bar\beta}(x,z_{\rm eq})
v^{\bar\gamma}_{\rm eq}\sigma^{;\bar\delta}(x,z_{\rm eq}) +O(y^3)
\,.
\end{eqnarray}
The bi-tensors necessary for the evaluation of
the direct force (\ref{eq:dir_force}) 
are $\sigma(x,\bar x)$ and ${\bar g}_{\alpha {\bar \alpha}}(x,\bar x)$, 
for which we consider the local coordinate expansion of these bi-tensors
around the coincidence limit $x\rightarrow {\bar x}$.
In a general regular coordinate system, 
$\sigma(x,{\bar x})$ and $\bar g_{\alpha\bar\alpha}(x,{\bar x})$
can be expanded as
\begin{eqnarray}
\sigma(x,{\bar x}) &=&
{1\over 2}g_{\alpha\beta}({\bar x})y^{\alpha\beta}
+\sum_{n=3,4,\cdots}{1\over n!}A_{\alpha^1\alpha^2\cdots\alpha^n}({\bar x})
y^{\alpha^1\alpha^2\cdots\alpha^n} \,, \label{eq:sigma-exp}
\\
{\bar g}_{\alpha{\bar\alpha}}(x,{\bar x}) &=&
g_{\alpha{\bar\alpha}}({\bar x})
+\sum_{n=1,2,\cdots}{1\over n!}
B_{\alpha{\bar\alpha}|\beta^1\beta^2\cdots\beta^n}({\bar x})
y^{\beta^1\beta^2\cdots\beta^n} \,; \label{eq:g-exp}
\\ 
y^{\alpha^1\alpha^2\cdots} &=& (x^{\alpha^1}-{\bar x}^{\alpha^1})
(x^{\alpha^2}-{\bar x}^{\alpha^2})\cdots.
\end{eqnarray}
To calculate the reaction force to a particle,
it is enough to know the expansion coefficients of $n=3,4$ of
 Eq.~(\ref{eq:sigma-exp}) 
and $n=1,2$ of Eq.~(\ref{eq:g-exp}).
 These are shown in Appendix A 
of Ref.~\cite{Mino:2001mq}. 
The local expansion of the force (\ref{eq:dir_force}) 
on the Boyer-Lindquist coordinates is quite tedious, though systematic. 
However, most of the terms give no contribution
to the harmonic coefficients in the coincidence limit $x \to z_0$.
Below, we shall focus on the terms that are non-vanishing
in the coincidence limit.

Without loss of generality, we may assume that the particle 
is located at $\theta_0=\pi/2$, $\phi_0=0$ at time $t_0$. 
Since the full force is calculated in the form of the Fourier-harmonic 
expansion and the Fourier modes are independent of the spherical 
harmonics, we may take the field point to lie on the hypersurface $t=t_0$ 
in the full force. Hence we may take $t=t_0$ before we perform 
the local coordinate expansion of the direct force. 
That is, we consider the local coordinate expansion of the direct force 
at a point $\{t_0,r,\theta,\phi\}$ near the particle localtion
 $\{t_0,r_0,\pi/2,0\}$. 

The local expansion of the direct force in the
Boyer-Lindquist coordinates can be done in such a way that
it consists of terms of the form,
\begin{eqnarray}
&& {R^{n_1}\Theta^{n_2}\phi^{n_3}\over \xi^{2n_4+1}}\,;
\label{eq:ingredient} \\
&& \qquad \xi:= \sqrt{2}r_0
\left(a-\cos\tilde\theta+{b\over 2}(\phi-\phi')^2\right)^{1/2}\,,
\\
\label{eq:xi}
&& \qquad R:=r-r_0\,,\quad\Theta:=\theta-{\pi\over2}\,, 
\end{eqnarray}
where $n_1$, $n_2$, $n_3$, $n_4$ are non-negative integers,
and $a$, $b$ and $\phi'$ defined by
\begin{eqnarray}
a &:=&
1+{1\over 2r_0^2}{r_0^2\over r_0^2+{\cal L}^2}
{r_0^2\over(r_0-2M)^2}{\cal E}^2R^2
\,, \\
b &:=& {{\cal L}^2\over r_0^2}
\,, \\
\phi'&:=&-{{\cal L}\over r_0^2+{\cal L}^2}u_rR \,,
\end{eqnarray}
where ${\cal E}:=-g_{tt}{dt/d\tau}$, ${\cal L}:=g_{\phi\phi}{d\phi/d\tau}$,
and $u_r:=g_{rr}{dr/d\tau}$,
and $\tilde\theta$ is the relative angle
between $(\theta,\phi)$ and $(\pi/2,\phi')$.

There are two apparently different terms in
the covariant form of the direct force given by Eq.~(\ref{eq:dir_force}); 
the first term in the curly brackets exhibiting the quadratic divergence, 
and the second term proportional to the curvature tensor that 
appears to be finite in the coincidence limit. 
In the local coordinate expansion, the second term will give terms
of the form $R/\xi$ or $\phi/\xi$. 
As discussed in Section~\ref{subsec:comp},
 the harmonic coefficients of $R/\xi$
vanish in the coincidence limit, while those of $\phi/\xi$
are finite but they give no contribution to the final result when
the infinite harmonic modes are summed up after
the coincidence limit is taken. 
Hence we may focus on the first term in the curly brackets of
Eq.~(\ref{eq:dir_force}).

 Since the orbit always remains 
on the equatorial plane, the force is symmetric under the 
transformation $\theta\to\pi-\theta$, which implies there is 
no term proportional to odd powers of $\Theta$. 
Hence we only need to consider the case of $n_2$ being an even 
number in the general form given by Eq.~(\ref{eq:ingredient}). 
Then the factor $\Theta^{n_2}$ 
may be eliminated by expressing $\Theta^2$ in terms of
$\xi$, $R$ and $\phi$, and we are left with terms of the form,
\begin{eqnarray}
{R^{n_1}\phi^{n_3} \over \xi^{2n_4+1}} \,.
\label{eq:dirform}
\end{eqnarray}
Explicitly, we find
\begin{eqnarray}
F_t^{\rm dir} &=& q\Biggl(
{\cal E} u_r{R\over\xi^3}+{\cal E}{\cal L}{\phi\over\xi^3}
\nonumber \\ && \qquad
-{1\over 2}{(r_0-2M){\cal E} u_r \over r_0^2}{1\over\xi}
+{2(r_0-2M){\cal E} {\cal L}^2 u_r \over r_0^2}{\phi^2\over\xi^3}
\nonumber \\ && \qquad
-{3\over 2}{(r_0-2M){\cal E} {\cal L}^4 u_r \over r_0^2}{\phi^4\over\xi^5}
\Biggr)
\,,
\label{eq:dirforcet}
\\
F_r^{\rm dir} &=& q\Biggl(
{{\cal L}^2\over r_0(r_0-2M)}{R\over\xi^3}
-{r_0^2{\cal E}^2\over (r_0-2M)^2}{R\over\xi^3}
-{\cal L}u_r{\phi\over\xi^3}
\nonumber \\ && \qquad
-{1\over 2}{2r_0^2+{\cal L}^2 \over r_0^3}{1\over\xi}
+{1\over 2}{{\cal E}^2 \over r_0-2M}{1\over\xi}
\nonumber \\ && \qquad
+{1\over 2}{(3r_0^2+4{\cal L}^2){\cal L}^2\over r_0^3}{\phi^2\over\xi^3}
-{2{\cal E}^2 {\cal L}^2 \over r_0-2M}{\phi^2\over\xi^3}
\nonumber \\ && \qquad
-{3\over 2}{(r_0^2+{\cal L}^2){\cal L}^4 \over r_0^3}{\phi^4\over\xi^5}
+{3\over 2}{{\cal E}^2 {\cal L}^4\over r_0-2M}{\phi^4\over\xi^5}
\,,
\label{eq:dirforcer}
\\
F_\theta^{\rm dir} &=& 0
\,,
\label{eq:dirforceth}
\\
F_\phi^{\rm dir} &=& q\Biggl(
-{\cal L} u_r {R\over\xi^3}-(r_0^2+{\cal L}^2){\phi\over\xi^3}
\nonumber \\ && \qquad
+{1\over 2}{(r_0-2M){\cal L} u_r \over r_0^2}{1\over\xi}
-{1\over 2}{(r_0-2M)(r_0^2+4{\cal L}^2){\cal L} u_r \over r_0^2}{\phi^2\over\xi^3}
\nonumber \\ && \qquad
+{3\over 2}{(r_0-2M)(r_0^2+{\cal L}^2){\cal L}^3 u_r \over r_0^2}
{\phi^4\over\xi^5}
\Biggr)
\,,
\label{eq:dirforceph}
\end{eqnarray}
where $F_{\alpha}^{\rm dir}=F_{\alpha}[\varphi^{\rm dir}]$.
The absence of $F_\theta^{\rm dir}$ is because of the
symmetry of the background. 

\subsection{Regularization counter terms}\label{subsec:comp}

What we have to do now is to perform the harmonic decomposition of the
components of the direct force given above.
To do so, we note the following important fact.
Apart from the trivial multiplicative factor of $R^{n_1}$ which
is independent of the spherical coordinates,
the terms to be expanded in the spherical harmonics are of
the form $\phi^{n_3}/\xi^{2n_4+1}$, or $(\phi-\phi')^{n_3}/\xi^{2n_4+1}$.
To the order of accuracy we need,
the factor $(\phi-\phi')^{n_3}$ may be eliminated by replacing
it to an equivalent $\phi$-derivative operator
of degree $n_3$ acting on $\xi^{2n_3-2n_4-1}$,
which is further converted to a polynomial in $m$ after
the harmonic expansion of $\xi^{2n_3-2n_4-1}$.
Thus the only basic formula we need is the harmonic expansion
of $\xi^{2p-1}$ where $p$ is an integer.
Note that, apart from the term $b(\phi-\phi')^2/2$ in $\xi$
with respect to which we expand $\xi$ in a convergent infinite series,
$\xi^{2p-1}$ is defined over the whole sphere to allow the straightforward
harmonic decomposition.
The result to the leading order in the coincidence limit $a \to1+0$ is
\begin{eqnarray}
\left({\xi\over\sqrt{2}\,r_0}\right)^{2p-1}
&=&\left(a-\cos\tilde\theta+{b\over 2}(\phi-\phi')^2\right)^{p-1/2}
\nonumber \\ 
&=& 2\pi\sum_{\ell=0}^{\infty}\sum_{m=-\ell}^{\ell}D^{p-1/2}_{\ell m}(a)
Y_{\ell m}(\theta,\phi)Y^*_{\ell m}(\theta',\phi')\,,
\label{eq:xipdec}
\\
D^{p-1/2}_{\ell m}(a) &\to&
 \cases{ \displaystyle
{1\over \sqrt{1+b}}{1\over -p-1/2}(a-1)^{p+1/2} \,, &for $p+{1\over2}<0$,
\cr \displaystyle
{(-1)^\ell 2^{p+1/2} \over \sqrt{1+b}} \sum_{n=0}^\infty K_n \,, 
& for $p+{1\over2}>0$\,; \cr} 
\nonumber \\ K_n &=& {\Gamma(p+1/2) \Gamma(p+n+1/2)\over
\Gamma(p+n-\ell+1/2)\Gamma(p+n+\ell+3/2)}{1\over n!}
\left({-m^2 b\over 1+b}\right)^n
\end{eqnarray}
We note that, although what we need here is only the case of an integer $p$,
the above formula is valid for any $p$ (except for the case $p=-1/2$).

After the decomposition, we can take the radial coincidence limit $r \to r_0$
(followed by the angular coincidence limit if desired). 
Here we briefly explain the reason why the terms proportional to
$R/\xi$ and $\phi/\xi$ give no contribution to the final
result. The term $R/\xi$ corresponds to $R$ times the case of $p=0$,
 for which $D_{\ell m}^{p-1/2}$ is finite in the limit $a\to1$
(i.e., $R\to0$). Hence all the coefficients vanish in the radial
coincidence limit.
As for $\phi/\xi$, it can be replaced by $(\phi-\phi')/\xi$
which is equivalent to $\partial_{\phi}\xi$ in the coincidence limit.
This corresponds to the case of $p=1$ multiplied by $m$.
Hence all the harmonic coefficients become odd functions of $m$,
and their sum over $m$ for each $\ell$ vanishes
in the angular coincidence limit.
As a result, the non-vanishing contribution comes only from
the terms $R/\xi^3$, $\phi/\xi^3$,
$1/\xi$, $\phi^2/\xi^3$ and $\phi^4/\xi^5$.

Barack and Ori define the regularization counter terms as 
\begin{eqnarray}\label{ABC}
\lim_{x\to z_0}F_{\alpha l}^{\rm dir}=A_{\alpha}L+B_{\alpha}
+C_{\alpha}/L+O(L^{-2}) \,. \\
D_{\alpha}=
\sum_{l=0}^{\infty}\left[\lim_{x\to z_0}F_{\alpha l}^{\rm dir}
-A_{\alpha}L-B_{\alpha}-C_{\alpha}/L\right].
\end{eqnarray}
where $F_{\alpha l}^{\rm dir}$ is the multipole $l$-mode 
of $F_{\alpha}^{\rm dir}$, 
$L=\ell+1/2$, and $A_{\alpha}$, $B_{\alpha}$ and $C_{\alpha}$
are independent of $L$.
The $A_\alpha$ term is to subract the quadratic divergence,
the $B_\alpha$ term the linear divergence, and the $C_\alpha$
term the logarithmic divergence. The $D_\alpha$ term is
the remaining finite contribution of the direct force to
be subtracted. We find $C_\alpha=D_\alpha=0$ in agreement with
Barack and Ori. We also find the complete agreement 
of $A_\alpha$ and $B_\alpha$ terms with their results 
for a general geodesic orbit as given below.
We call these feature of regularization counter terms 
{\it the standard form}. 

The $A$-term describes the quadratic divergent terms of the direct force.
It comes from the terms $R / \xi^{3}$ in Eqs.~(\ref{eq:dirforcet})
$\sim$ (\ref{eq:dirforceph}).
The important fact is that it is odd in $R$.
This leads to the harmonic coefficients proportional to
 ${\rm sign}(R)$.
We find
\begin{eqnarray}
A_{t} &=& {\rm sign}(R) {q^2 \over r_0^2} {r_0-2M \over r_0}
{u_r \over 1+{\cal L}^2/r_0^2} \,,
\\
A_{r} &=& -{\rm sign}(R) {q^2 \over r_0^2} {r_0 \over r_0-2M}
{{\cal E} \over 1+{\cal L}^2/r_0^2} \,,
\\
A_{\phi} &=& 0 \,.
\label{eq:Aterm}
\end{eqnarray}
These $A$-terms vanish when averaged over both limits $R\to\pm0$.

The $B$-term describes linearly divergent terms, which
are of the form, $\phi^{2n} / \xi^{2n+1}$ 
in Eqs.~(\ref{eq:dirforcet}) $\sim$ (\ref{eq:dirforceph}).
We find the $B$-term in terms of the hypergeometric functions as 
\begin{eqnarray}
B_t &=& -{(r_0-2M){\cal E}u_r \over 2r_0^3}
F\left({3\over2},{3\over2};1;-{{\cal L}^2\over r_0^2}\right) \,,
\\
B_r &=& {(r_0-2M)u_r^2 \over 2r_0^3}
F\left({3\over2},{3\over2};1;-{{\cal L}^2\over r_0^2}\right)
\nonumber \\ && -{1\over 2r_0^2}
\left(
F\left({1\over2},{1\over2};1;-{{\cal L}^2\over r_0^2}\right)
+{{\cal L}^2 \over 2r_0^2}
F\left({3\over2},{3\over2};2;-{{\cal L}^2\over r_0^2}\right)
\right) \,,
\\
B_{\phi} &=& {(r_0-2M){\cal L}u_r\over 16r_0^3}
\left(
8F\left({3\over2},{3\over2};1;-{{\cal L}^2\over r_0^2}\right) \right.
\nonumber \\ && \left.
-4F\left({3\over2},{3\over2};2;-{{\cal L}^2\over r_0^2}\right)
+{9{\cal L}^2 \over r_0^2}
F\left({5\over2},{5\over2};3;-{{\cal L}^2\over r_0^2}\right)
\right) \,.
\label{eq:Bterm}
\end{eqnarray}
As mentioned at the beginning of this section, the above results for the 
$A$ and $B$-terms perfectly agree with
the results obtained by Barack and Ori in a quite different
fashion~\cite{Barack:2002mh,Barack:2002bt,Barack:2001gx}.

\section{Gravitational Self-force in the Regge-Wheeler gauge}\label{sec:gauge}

Recently, Detweiler and Whiting found a slight but important
modification of the direct and tail parts of the metric 
perturbation~\cite{Detweiler:2002mi}.
The full metric perturbation in the harmonic gauge
is now decomposed as
\begin{eqnarray}
h^{\rm full,H}_{\mu\nu}=h^{\rm S,H}_{\mu\nu}+h^{\rm R,H}_{\mu\nu}\,,
\end{eqnarray}
where the superscript H stands for the harmonic gauge,
and the S part, $h^{\rm S,H}_{\mu\nu}$, satisfies the same
linearized Einstein equations with source as $h^{\rm S,full}_{\mu\nu}$,
\begin{eqnarray}
\Box\,\bar{h}^{{\rm S,H}}_{\mu\nu}
+2R_{\mu}{}^{\alpha}{}_{\nu}{}^{\beta}\bar{h}^{{\rm S,H}}_{\alpha\beta}
=-16\pi T_{\mu\nu} \,,
\end{eqnarray}
where 
$\bar{h}_{\alpha\beta}
=h_{\alpha\beta}-\frac{1}{2}g_{\alpha\beta}h_{\mu}{}^{\mu}$.
The new tail part, called the R part, $h^{\rm R,H}_{\mu\nu}$,
is then a homogeneous solution. Detweiler and Whiting showed
that the properly regularized self-force is given by
the R part of the metric perturbation.
Namely, we have
\begin{eqnarray}
\frac{d^2 z^{\alpha}}{d\tau^2}
+\Gamma_{\mu\nu}^{\alpha}\frac{dz^{\mu}}{d\tau}\frac{dz^{\nu}}{d\tau}
={1 \over \mu}\,F^{\alpha}[h^{\rm R,H}] \,,
\end{eqnarray}
where $z^{\alpha}(\tau)$ is an orbit of the particle
parametrized by the background proper time
 (i.e., $g_{\mu\nu} (dz^\mu/d\tau)(dz^\nu/d\tau)=-1$),
and 
\begin{eqnarray}
F_{\alpha}[h] =
-\mu P_{\alpha}{}^{\beta}
(\bar{h}_{\beta\gamma;\delta}
-\frac{1}{2}g_{\beta\gamma}
 {\bar{h}^{\epsilon}}{}_{\epsilon;\delta}
-\frac{1}{2}\bar{h}_{\gamma\delta;\beta}
+\frac{1}{4}g_{\gamma\delta}
{\bar{h}^{\epsilon}}{}_{\epsilon;\beta}
)u^{\gamma}u^{\delta} \,,
\label{eq:formalF}
\end{eqnarray}
where ${P_{\alpha}}^{\beta}={\delta_{\alpha}}^{\beta}+u_{\alpha}u^{\beta}$ 
and $u^{\alpha}=dz^\alpha/d\tau$.

Since the R part is a solution of the Einstein equations,
it is perfectly legitimate to consider a gauge transformation
of it to another gauge. Therefore we can {\it define} the
R part in an arbitrary gauge by a gauge transformation of
the R part in the harmonic gauge to that gauge.
We thus have an unambiguous definition of the self-force in
an arbitrary gauge. In particular, we may 
consider the self-force in the Regge-Wheeler (RW) gauge, in which
we may be able to obtain the full metric perturbation.

In this section, we formulate a method to obtain the self-force in 
the RW gauge by applying the mode-decomposition regularization.
Then, as a simple but non-trivial example, we consider a
circular orbit and derive the regularized self-force 
in the RW gauge to 1PN order~\cite{Nakano:2003he}.

\subsection{Gauge transformation to the RW gauge}

The gravitational self-force 
acting on the particle is given by the R part in the harmonic gauge, 
formally obtained as
\begin{eqnarray}
F_\alpha^{\rm H}(\tau) &=& \lim_{x \to z(\tau)}
F_\alpha \left[h^{\rm R,H}\right](x)
\nonumber \\
&=&
\lim_{x \to z(\tau)}
F_\alpha \left[h^{\rm full,H}-h^{\rm S,H} \right](x)
\nonumber \\
&=&
\lim_{x \to z(\tau)}
\left(F_\alpha \left[h^{\rm full,H} \right](x)
-F_\alpha \left[h^{\rm S,H} \right](x)
\right)
\,, \label{eq:force2}
\end{eqnarray}
The S part can be calculated by the method 
discussed in the previous section. It is important to note that
the S and R parts independently satisfy the harmonic gauge condition.

We consider the gauge transformation of the R part
from the harmonic gauge to the RW gauge,
\begin{eqnarray}
x^{{\rm H}}_{\mu} &\to&
x^{{\rm RW}}_{\mu} = x^{{\rm H}}_{\mu}
+ \xi_{\mu}^{{\rm H} \to {\rm RW}} \,,
\label{displace} \\
h_{\mu\nu}^{{\rm R,H}} &\to&
h_{\mu\nu}^{{\rm R,RW}} =
h_{\mu\nu}^{{\rm R,H}} - 2\,
\nabla \xi^{{\rm H}\to{\rm RW}}_{(\mu;\nu)}\,,
\label{gauge-trans}
\end{eqnarray}
where $\xi_{\mu}^{{\rm H} \to {\rm RW}}$
is the generator of the gauge transformation.
Then the self-force in the RW gauge is
given by
\begin{eqnarray}
F_\alpha^{\rm RW}(\tau)
&=&\lim_{x \to z(\tau)}
F_\alpha\left[h^{\rm R,RW}\right]
\nonumber \\ 
&=& \lim_{x \to z(\tau)}
F_\alpha \left[h^{\rm R,H} - 2\,\nabla\xi^{{\rm H} \to {\rm RW}}
\left[h^{\rm R,H}\right] \right](x)
\nonumber\\
&=&
\lim_{x \to z(\tau)}
F_\alpha \left[h^{\rm full,H}-h^{\rm S,H}
- 2\,\nabla\xi^{{\rm H} \to {\rm RW}}
\left[h^{\rm full,H}-h^{\rm S,H}\right]
\right](x)
\nonumber \\
&=&
\lim_{x \to z(\tau)}
F_\alpha \left[h^{\rm full,H}
- 2\,\nabla\xi^{{\rm H} \to {\rm RW}}
\left[h^{\rm full,H}\right]
\right. \nonumber \\ && \left. \qquad \qquad 
-h^{\rm S,H}
+ 2\,\nabla\xi^{{\rm H} \to {\rm RW}}
\left[h^{\rm S,H}\right]
\right](x)
\nonumber \\
&=&
\lim_{x \to z(\tau)}
\Bigl(
F_\alpha \left[h^{\rm full,RW}\right](x)
\nonumber \\ && \qquad \qquad 
-F_\alpha \left[h^{\rm S,H}-2\,\nabla\xi^{{\rm H} \to {\rm RW}}
\left[h^{\rm S,H}\right]
\right](x)
\Bigr)\,,
\label{eq:RWsf}
\end{eqnarray}
where we have omitted the spacetime indices of $h_{\mu\nu}$
and $\nabla_{(\mu}\xi_{\nu)}$ for notational simplicity.
The full metric perturbation $h^{\rm full,RW}_{\mu\nu}$
can be calculated by using the Regge-Wheeler-Zerilli formalism,
while the S part $h^{\rm S,H}_{\mu\nu}$ can be obtained with
sufficient accuracy by the local analysis near the particle
location. Thus provided that the gauge transformation of
the S part can be unambiguously defined, we will be able to
obtain the regularized self-force in the RW gauge.
Namely, the following gauge transformation should be 
well defined:
\begin{eqnarray}
\xi_\alpha^{{\rm S},{\rm H} \to {\rm RW}}
\equiv\xi_\alpha^{{\rm H} \to {\rm RW}}\left[h_{\mu\nu}^{\rm S,H}\right]\,.
\label{eq:Sgtrans}
\end{eqnarray}

Note that the self-force (\ref{eq:RWsf}) is almost identical
to the expression obtained in the intermediate
gauge approach~\cite{Barack:2001ph}, if we replace
the S and R parts by the direct and tail parts,
respectively. The only difference is that the S and R parts are
now solutions of the inhomogeneous and homogeneous Einstein
equations, respectively. Hence the S part in the
RW gauge is (at least formally) well-defined provided that
the gauge transformation of the S part, Eq.~(\ref{eq:Sgtrans})
is unique. As will be shown later in
Eqs.~(\ref{eq:GGT4}),
this turns out to be indeed the case.
Therefore one may identify
the self-force (\ref{eq:RWsf}) to be actually the one evaluated
in the RW gauge~\cite{Minopc}, not in some intermediate gauge.

\subsection{Full force for circular orbits}
\label{sec:full}

We consider the full metric perturbation 
and its self-force in the case of a circular orbit.
Hereafter, we will need to consider the case of general,
non-circular orbits. However, because of several technical problems,
which are not particular to the gravitational self-force
but common to all types of the self-force, we leave it for 
future work. A promising method to deal with general, non-circular
orbits will be presented later in Sec.~\ref{seq:newreg}, in which
the case of scalar self-force is discussed for simplicity.
An extension to the gravitational case is under progress~\cite{future}.

First, the metric perturbation is calculated 
by the Regge-Wheeler-Zerilli formalism in which the metric
is expanded in terms of Fourier series in frequency $\omega$
and tensor spherical harmonics with eigenvalue indices $(\ell, m)$
by using the symmetry of the background spacetime.
Then, we derive the Fourier-harmonic components of
the self-force. As noted earlier, although the full self-force
diverges on the orbit of the particle, each Fourier-harmonic component
is finite. We may even sum over $\omega$ and $m$ and the
result is still finite. Thus we formally express the full self-force
in the coincidence limit $x\to z(\tau)$ as
\begin{eqnarray}
F^\alpha[h^{\rm full}]=\sum_{\ell=0}^\infty F^\alpha_\ell(r_0) \,,
\end{eqnarray}
where $r=r_0$ is the radius of a circular orbit.

On the Schwarzschild background, 
the metric perturbation $h_{\mu\nu}$ can be expanded in terms of
tensor harmonics as
\begin{eqnarray}
\mathbf{h} &=& \sum_{\ell m} \left[
f(r)H_{0\ell m}(t,r)\mathbf{a}^{(0)}_{\ell m}
-i\sqrt{2}H_{1\ell m}(t,r)\mathbf{a}^{(1)}_{\ell m}
+\frac{1}{f(r)}H_{2\ell m}(t,r)\mathbf{a}_{\ell m}
\right. \nonumber \\
& & \quad 
-\frac{i}{r}\sqrt{2\ell(\ell+1)}h^{(e)}_{0\ell m}(t,r)\mathbf{b}^{(0)}_{\ell m}
+\frac{1}{r}\sqrt{2\ell(\ell+1)}h^{(e)}_{1\ell m}(t,r)\mathbf{b}_{\ell m}
\nonumber \\
& & \quad 
+\sqrt{\frac{1}{2}\ell(\ell+1)(\ell-1)(\ell+2)}G_{\ell m}(t,r)\mathbf{f}_{\ell m}
\nonumber \\ & & \quad 
+\left(\sqrt{2}K_{\ell m}(t,r)
 -\frac{\ell(\ell+1)}{{\sqrt{2}}}G_{\ell m}(t,r)\right)\mathbf{g}_{\ell m}
\nonumber \\
& & \quad 
-\frac{\sqrt{2\ell(\ell+1)}}{r}h_{0\ell m}(t,r)\mathbf{c}^{(0)}_{\ell m}
+\frac{i\sqrt{2\ell(\ell+1)}}{r}h_{1\ell m}(t,r)\mathbf{c}_{\ell m}
\nonumber \\ & & \left. \quad
+\frac{\sqrt{2\ell(\ell+1)(\ell-1)(\ell+2)}}{2r^2}h_{2\ell m}(t,r)\mathbf{d}_{\ell m}
\right] \,, \label{eq:hharm}
\end{eqnarray}
where $\mathbf{a}_{\ell m}^{(0)}$, $\mathbf{a}_{\ell m},~\cdots$ are the
tensor harmonics introduced by Zerilli~\cite{Zer}.
The energy-momentum tensor of a point particle takes the form,
\begin{eqnarray}
T^{\mu\nu}&=& \mu \int^{+\infty}_{-\infty} \delta^{(4)}(x-z(\tau))
{dz^{\mu} \over d\tau}{dz^{\nu} \over d\tau}d\tau \nonumber \\
&=& \mu {1 \over u^t}u^{\mu}u^{\nu}
{\delta(r-r_0(t)) \over r^2} \delta ^{(2)}
({\bf\Omega}-{\bf\Omega}_0(t)) \,.
\end{eqnarray}
The RW gauge is defined by the conditions on the
metric perturbation as
\begin{eqnarray}
h_2^{\rm RW}
=h_{0}^{(e){\rm RW}}=h_{1}^{(e){\rm RW}}=G^{\rm RW}=0\,.
\label{eq:RWcond}
\end{eqnarray}
The Regge-Wheeler and Zerilli equations are obtained
by plugging the metric perturbation (\ref{eq:hharm}) into
the linearized Einstein equations and expand it in the
Fourier series. The Regge-Wheeler-Zerilli formalism has been
improved recently by Jhingan and Tanaka~\cite{Jhingan:2002kb}.
Here, however, we use the original formalism developed by
Regge and Wheeler, and by Zerilli.
The tensor harmonics can be further classified into even and
odd parities. Even parity modes are defined by the parity
$(-1)^\ell$ under the transformation $(\theta,\phi)\to(\pi-\theta,\phi+\pi)$,
while odd parity modes are by the parity $(-1)^{\ell+1}$.

For odd parity modes, we introduce a new radial function
$R_{\ell m \omega}^{{\rm (odd)}}(r)$
in terms of which the two radial functions of the metric perturbation
are expressed as (also see \cite{Nakano:2003he} 
for $h_{0\ell m \omega}^{\rm RW}$)
\begin{eqnarray}
h_{1\ell m \omega}^{\rm RW}
&=&{r^2  \over (r-2M)}R_{\ell m \omega}^{{\rm (odd)}} \,.
\label{eq:h0}
\end{eqnarray}
The new radial function $R_{\ell m \omega}^{{\rm (odd)}}(r)$ 
satisfies the Regge-Wheeler equation,
\begin{eqnarray}
&& {d^2 \over dr^{*2}} R_{\ell m \omega}^{{\rm (odd)}}
+[\omega ^2 -V_{\ell}(r)]
R_{\ell m \omega}^{{\rm (odd)}} 
={8\pi i \over [{1\over2}\ell(\ell+1)(\ell-1)(\ell+2)]^{1/2}}{r-2M \over r^2}
\nonumber \\ && \quad
\times \left(-r^2{d \over dr}[(1-{2M \over r})D_{\ell m \omega}]
+(r-2M)[(\ell-1)(\ell+2)]^{1/2}Q_{\ell m \omega} \right) \,,
\end{eqnarray}
where
$r^*=r+2M \log (r/2M-1)$, and the potential $V_{\ell}$ is
given by
\begin{eqnarray}
V_{\ell}(r)
=\left( 1-{2M \over r} \right)
\left( {\ell(\ell+1) \over r^2}-{6M \over r^3} \right) \,.
\end{eqnarray}
The source term $Q_{\ell m \omega}$ vanishes
in the case of a circular orbit and
\begin{eqnarray}
D_{\ell m\omega}(r) &=& 
\left[\frac{\ell(\ell+1)(\ell-1)(\ell+2)}{2}\right]^{-1/2}
\mu\,{(u^{\phi})^2\over u^t}
\nonumber \\ && \times
\delta(r-r_0)\,m\,\partial_{\theta}\,Y_{\ell m}^*(\theta_0,\,\phi_0) \,,
\end{eqnarray}
where the orbit is given by
\begin{eqnarray}
z^\alpha(\tau)=
\left\{u^t \tau ,\, r_0 ,\, {\pi\over 2} ,\, u^\phi \tau \right\}
\,;\,
u^t = \sqrt{r_0 \over r_0-3M} \,, \,
u^\phi = {1\over r_0}\sqrt{M\over r_0-3M} \,,
\label{eq:orbit}
\end{eqnarray}
where $\Omega=u^\phi/u^t=\sqrt{M/r_0^3}$ which is the orbital frequency.
The orbit is assumed to be on the equatorial plane
without loss of generality.

For even parity modes with the parity $(-1)^{\ell}$,
we introduce a new radial function $R_{\ell m \omega}^{{\rm (Z)}}(r)$
in terms of which the four radial functions
of the metric perturbation are expressed as 
(also see \cite{Nakano:2003he} for 
$H_{0\ell m \omega}^{\rm RW},\,H_{1\ell m \omega}^{\rm RW}$ and 
$H_{2\ell m \omega}^{\rm RW}$)
\begin{eqnarray}
K_{\ell m \omega}^{\rm RW}
&=&{\lambda (\lambda +1)r^2+3\lambda Mr+6M^2 \over r^2(\lambda r+3M)}
R_{\ell m \omega}^{{\rm (Z)}}
+{r-2M \over r}{d  \over dr} R_{\ell m \omega}^{{\rm (Z)}}
\cr
&& -{r(r-2M) \over \lambda r+3M}\tilde C_{1\ell m \omega}
+{i (r-2M)^2 \over r(\lambda r+3M)}\tilde C_{2\ell m \omega} \,, 
\label{eq:H2}
\end{eqnarray} 
where $\lambda = (\ell-1)(\ell+2)/2$ and the local source terms are given by
\begin{eqnarray}
\tilde B_{\ell m \omega}
&=&{8\pi r^2(r-2M) \over \lambda r+3M}\{A_{\ell m \omega}
+[{1\over2}\ell(\ell+1)]^{-1/2}B_{\ell m \omega}\}
\nonumber \\ && 
-{4\pi \sqrt{2} \over \lambda r +3M}
{Mr \over \omega}A_{\ell m \omega}^{(1)} \,, \cr
\tilde C_{1\ell m \omega}&=&{8\pi \over \sqrt{2}\omega}A_{\ell m \omega}^{(1)}
+{1 \over r}\tilde B_{\ell m \omega}
-16\pi r [{1 \over 2}\ell(\ell+1)(\ell-1)(\ell+2)]^{-1/2}F_{\ell m \omega} \,,
\cr
\tilde C_{2\ell m \omega}&=&-{8\pi r^2 \over i \omega}
{[{1\over2}l(l+1)]^{-1/2} \over r-2M}B_{\ell m \omega}^{(0)}
-{ir \over r-2M}\tilde B_{\ell m \omega}
\cr
&& +{16\pi ir^3 \over r-2M}
[{1 \over 2}\ell(\ell+1)(\ell-1)(\ell+2)]^{-1/2}F_{\ell m \omega}
\,.
\end{eqnarray}
Here the harmonic coefficients of the source terms $A_{\ell m \omega}$, 
$A_{\ell m \omega}^{(1)}$ and
$B_{\ell m \omega}$ vanish in the circular case and 
\begin{eqnarray}
B_{\ell m \omega}^{(0)}&=&
\left[{\ell(\ell+1)\over 2}\right]^{-1/2}
\mu \,u^{\phi} \left(1-{2M\over r}\right){1\over r}
\delta(r-r_0)\,m\,Y_{\ell m}^*(\theta_0,\,\phi_0)
\,,
\cr
F_{\ell m \omega}&=&
{1\over 2}\left[{\ell(\ell+1)(\ell-1)(\ell+2)\over 2}\right]^{-1/2}
\mu \,{(u^{\phi})^2  \over u^t}
\nonumber \\ && \times 
\delta(r-r_0)\,(\ell(\ell+1)-2m^2)\,Y_{\ell m}^*(\theta_0,\,\phi_0)
\,.
\end{eqnarray}
The new radial function $R_{\ell m \omega}^{{\rm (Z)}}(r)$
obeys the Zerilli equation,
\begin{eqnarray}
{d^2 \over dr^{*2}} R_{\ell m \omega}^{{\rm (Z)}}
+[\omega ^2 -V_{\ell}^{{\rm (Z)}}(r)]R_{\ell m \omega}^{{\rm (Z)}}
=S_{\ell m \omega}^{{\rm (Z)}} \,,
\label{eq:elederi}
\end{eqnarray}
where
\begin{eqnarray}
V_{\ell}^{{\rm (Z)}}(r)=\left( 1-{2M \over r} \right)
{2\lambda ^2(\lambda +1)r^3+6\lambda ^2 Mr^2+18\lambda M^2r+18M^3
\over r^3(\lambda r+3M)^2} \,,
\end{eqnarray}
and
\begin{eqnarray}
S_{\ell m \omega}^{{\rm (Z)}}&=&-i{r-2M \over r}{d \over dr}
\left[{(r-2M)^2 \over r(\lambda r+3M)}\left({ir^2 \over r-2M}
\tilde C_{1\ell m \omega}
+\tilde C_{2\ell m \omega} \right)\right]
\nonumber \\ &&
+i{(r-2M)^2 \over r(\lambda r+3M)^2}
\left[{\lambda (\lambda +1)r^2+3\lambda Mr+6M^2 \over r^2}
\tilde C_{2\ell m \omega}
\right. \nonumber \\ && \left. 
+i{\lambda r^2-3\lambda Mr-3M^2 \over (r-2M)}
\tilde C_{1\ell m \omega}\right] \,.
\end{eqnarray}
The Zerilli equation can be transformed to the Regge-Wheeler equation
by the Chandrasekhar transformation \cite{chandra} if desired.

The homogeneous solutions of the Regge-Wheeler equation are discussed
in detail by Mano {\it et al.} \cite{Mano2}.
By constructing the retarded Green function from the homogeneous
solutions with appropriate boundary conditions, namely, the two
independent solutions with the in-going and up-going wave boundary
conditions, we can solve the Regge-Wheeler and Zerilli equations to
obtain the full metric perturbation in the RW gauge.
Here, we consider the radial functions up to the first post-Newtonian
(1PN) order. 

The radial function for the odd part of the 
metric perturbation is obtained for $r<r_0$ as
\begin{eqnarray}
R_{\ell m \omega}^{{\rm (odd)}}(r) =
{\displaystyle \frac {16\,i\,\pi \,\mu \,\Omega ^{2}\,m\,r}
{(2\,\ell + 1)\,\ell\,(\ell + 1)\,(\ell + 2)}}
\,\left( {\displaystyle \frac {r}{r_0}} \right)^{\ell}
\partial_{\theta}Y_{\ell m}^*(\theta_0,\,\phi_0) \,, 
\label{eq:Rodd}
\end{eqnarray}
where $\Omega=u^{\phi}/u^t$ (see \cite{Nakano:2003he} for $r>r_0$). 
For the even part, the radial function
is obtained for $r<r_0$ (also see \cite{Nakano:2003he} for $r>r_0$) as
\begin{eqnarray}
R_{\ell m \omega}^{{\rm (Z)}} &=&
{8\, \Omega \,m\,\pi \,u^t\,\mu
\over (2\,\ell + 1)\,(\ell + 2)\,(\ell + 1)
\omega }
\biggl( 2{r \over r_0} 
+ 2\,{\displaystyle \frac {(\ell^{2} - 2\,\ell - 1)\,M\,r}{(\ell - 1)\,r_0^2}}
\nonumber \\ && \quad 
+ \left( - \,{\displaystyle \frac {r^{3}}{(2\,\ell + 3)\,r_0}}  +
{\displaystyle \frac {(\ell^{2} - \ell + 4)\,r_0\,r}
{\ell\,(2\,\ell - 1)\,(\ell - 1)}} \right) \,\omega ^{2}
\nonumber \\ && \quad
- {\displaystyle \frac {2\,(\ell^{4} + \ell^{3} - 6\,\ell^{2}
- 4\,\ell - 4)\,M}{\ell\,(\ell - 1)\,(\ell + 2)\,r_0}} \biggr)
\,\left( {\displaystyle \frac {r}{r_0}} \right)^{\ell}
Y_{\ell m}^*(\theta_0,\,\phi_0) \,.
\label{eq:Reven}
\end{eqnarray}
The metric perturbation in the RW gauge is obtained from
Eqs.~(\ref{eq:h0}) and (\ref{eq:H2}).

Next, we calculate the self-force. We apply the
`mode-decomposition regularization' method, in which
the force is decomposed into harmonic modes and
subtract the harmonic-decomposed S part mode by mode before
the coincidence limit $x\to z(\tau)$ is taken. 
Since the orbit under consideration is circular, the source term
contains the factor $\delta(\omega-m\Omega)$, and the frequency integral
can be trivially performed. Hence we can calculate the harmonic coefficients
of the full metric perturbation in the time-domain. This is a great
advantage of the circular orbit case, since the S part can be given
only in the time-domain.
We also note that the $\theta$-component of the force vanishes because
of the symmetry, and $F^{\phi}=[(r_0-2M)/(r_0^3 \Omega)] F^t$
for a circular orbit. 

Performing the summation over $m$, we find in the end,
\begin{eqnarray}
\left. F^t_{{\rm full,RW}}\right|_{\ell}
&=& \left. F^{\theta}_{{\rm full,RW}}\right|_{\ell} 
= \left. F^{\phi}_{{\rm full,RW}}\right|_{\ell} = 0 \,, \cr
\left. F^{r(+)}_{{\rm full,RW}}\right|_{\ell} &=&
- {\displaystyle \frac {(\ell + 1)\,\mu ^{2}}{r_0^{2}}}
+ {\displaystyle \frac {1}{2}} \,
{\displaystyle 
\frac {\mu ^{2}\,(12\,\ell^{3} + 25\,\ell^{2} + 4\,\ell - 21)\,M}
{r_0^{3}\,(2\,\ell + 3)\,(2\,\ell - 1)}}
\,, \cr
\left. F^{r(-)}_{{\rm full,RW}}\right|_{\ell} &=&
 {\displaystyle \frac {\ell\,\mu ^{2}}{r_0^{2}}}
- {\displaystyle \frac {1}{2}}
\,{\displaystyle \frac {\mu ^{2}\,(12\,\ell^{3} + 11\,\ell^{2} - 10\,\ell + 12)\,M}
{(2\,\ell - 1)\,(2\,\ell + 3)\,r_0^{3}}}
\,.
\end{eqnarray}
We see that the only non-vanishing component is the radial component
as expected because there is no radiation reaction effect at 1PN order.
In the above, the index $(+)$ denotes that the coincidence limit
is taken from outside ($r>r_0)$ of the orbit, and $(-)$ from
inside ($r<r_0$) of the orbit, and the vertical bar with $\ell$, 
${\cdots}~\Bigl|_{\ell}$, 
denotes the coefficient of the $\ell$ mode in the coincidence limit. 
We note that the above result is valid for $\ell \geq 2$. 
There are some complications with $\ell=0$ and $1$ modes,
and they need to be treated separately. The details are
discussed in \cite{Nakano:2003he,DP}. Here we focus on
the modes $\ell\geq2$.

\subsection{The S part in the harmonic gauge}
\label{sec:dir}

In this subsection, we present the S part of the metric perturbation
and its self-force by using the local coordinate expansion.
It may be noted that if we simply apply the scalar harmonic expansion
to each component of the self-force, we obtain the exactly same
regularization counter terms as in the case of a scalar charge
discussed in Sec.~\ref{sec:MDR}.
However, here we consider the tensor harmonic expansion.
So, each $\ell$ component may be different from the scalar case.

The S part of the metric perturbation in the harmonic gauge
is given covariantly as
\begin{eqnarray}
&& \bar{h}_{\mu\nu}^{\rm S,H}(x) = 4 \mu \left[{
\bar g_{\mu\alpha}(x,z_{\rm ret})\bar g_{\nu\beta}(x,z_{\rm ret})
u^\alpha(\tau_{\rm ret})u^\beta(\tau_{\rm ret})
\over \sigma_{;\gamma}(x,z_{\rm ret})u^\gamma(\tau_{\rm ret})}
\right]
\nonumber \\ && 
+ 2 \mu (\tau_{{\rm adv}}-\tau_{{\rm ret}})
\bar g_{\mu}{}^{\alpha}(x,z_{\rm ret})
\bar g_{\nu}{}^{\beta}(x,z_{\rm ret})
R_{\gamma \alpha \delta \beta}(z_{\rm ret})
u^\gamma(\tau_{{\rm ret}})u^\delta(\tau_{{\rm ret}}) 
\nonumber \\ && 
+O(y^2) \,,
\label{eq:spart}
\end{eqnarray}
where
$\bar h_{\mu\nu}=h_{\mu\nu}-1/2\,g_{\mu\nu}h_{\alpha}{}^{\alpha}$ is the trace-reversed
metric perturbation, $z_{\rm ret}=z(\tau_{\rm ret})$, $\tau_{\rm ret}$ 
and $\tau_{\rm adv}$ are the retarded proper time and advanced 
proper time, respectively, defined by the cross section of the
past and future light cones of the field point $x$
with the orbit. As in Sec.~\ref{sec:MDR},
$\bar g_{\mu\alpha}$ is the parallel displacement bi-vector, and
$y$ is the difference of the coordinates between
$x$ and $z_0$, $y^\mu=x^\mu-z_0^\mu$.
Details of the local expansion are given in Ref.~\cite{Mino:2001mq}.
The difference between the S part and the direct part
appears in the terms of $O(y)$, i.e., the second term on the
right-hand side of Eq.~(\ref{eq:spart}). It turns out, however, that
this term does not contribute to the self-force.

The local coordinate expansion of the S part is then
performed in exactly the same manner as in Sec.~\ref{sec:MDR},
with a suitable extension of local quantities over the whole
sphere. It is in principle possible to calculate the harmonic 
coefficients of the extended S part exactly. 
However, it is neither necessary nor meaningful because 
the extended S part is only approximate. 
In fact, corresponding to the fact that all the terms in positive
powers of $y$ vanish in the coincidence limit, it is known 
that all the terms of $O(1/L^2)$ or higher,
where $L=\ell+1/2$, vanish when summed over $\ell$ \cite{Mino:2001mq}
in the harmonic gauge. It should be noted, however, this result 
is obtained by expanding the force in the 
scalar spherical harmonics. In our present analysis, we employ
the tensor spherical harmonic expansion. So, the meaning of
the index $\ell$ is different. 
Nevertheless, the same is found to be true.
 Namely, by expanding the S part of
the metric perturbation in the tensor spherical harmonics,
the S force in the harmonic gauge is found to have the form,
\begin{eqnarray}
\left. F^{\mu(\pm)}_{\rm S,H}\right|_\ell=\pm A^\mu L+B^\mu +D^\mu_\ell\,,
\label{eq:S-force}
\end{eqnarray}
where $A^\mu$ and $B^\mu$ are independent of $\ell$,
and the $\pm$ denotes that the limit to $r_0$ is taken 
from the greater or smaller value of $r$, and
\begin{eqnarray}
D^\mu_\ell=\frac{d^\mu}{L^2-1}+
\frac{e^\mu}{(L^2-1)(L^2-4)}+
\frac{f^\mu}{(L^2-1)(L^2-4)(L^2-9)}+
\cdots.
\end{eqnarray}
Then, the summation of $D^\mu_\ell$ over $\ell$ 
(from $\ell=0$ to $\infty$) vanishes.
For convenience, let us call this the standard form.
As we shall see later, the standard form of the S force 
is found to persist also in the RW gauge, at least in
the present case of a circular orbit to 1PN order. 

For the moment, let us assume the standard form of the
S force both in the harmonic gauge and the RW gauge.
Then, we may focus our discussion on the divergent terms.
When we calculate the S~force in the RW gauge,
we first transform the metric perturbation from the harmonic
gauge to the RW gauge, and then take appropriate linear combinations
of their first derivatives. We then find that
the harmonic coefficients 
$h_{2\ell m}^{\rm S,H}$, $h_{0\ell m}^{\rm (e) S,H}$ and
$h_{1\ell m}^{\rm (e) S,H}$
are differentiated two times, and $G_{\ell m}^{\rm S,H}$
is differentiated three times,
while the rest are differentiated once, to obtain the S force.
So, it is necessary and sufficient to perform the Taylor expansion
of the harmonic coefficients up to $O(X^2)$ for
$h_{2\ell m}^{\rm S,H}$, $h_{0\ell m}^{\rm (e) S,H}$ and
$h_{1\ell m}^{\rm (e) S,H}$,
and up to $O(X^3)$ for $G_{\ell m}^{\rm S,H}$,
and the rest up to $O(X)$, where $X=T(\equiv t-t_0)$ or $R(\equiv r-r_0)$.
To the accuracy mentioned above, 
the harmonic coefficients of the S part are found in the form, for example, 
\begin{eqnarray}
h_{0\ell m}^{\rm S,H}(t,\,r)
&=& {2 \over L}\,\pi\, \mu \biggl[
\frac {4\,i\,T\,m\,
r_0\,(L^{2} - 2)\,(u^{\phi})^2}
{{\cal L}^{(2)}\,(L^2 - 1)} 
 +\cdots\,\biggr]\partial_{\theta}\,Y_{\ell m}^*(\theta_0,\,\phi_0) 
\,,
\label{eq:harmonicS}
\end{eqnarray}
where we have defined
\begin{eqnarray}
{\cal L}^{(2)} &=& \ell(\ell+1) = \left(L^2-{1\over 4}\right) \,.
\end{eqnarray}
The above example is the coefficient obtained by approaching
the orbit from inside ($r<r_0$).
The other coefficients, as well as the coefficients obtained
in the case of approaching to the orbit from outside ($r>r_0$),
can be found on the web page: 
{http://www2.yukawa.kyoto-u.ac.jp/\~{}misao/BHPC/}. 

Now we consider the S force in the harmonic gauge. 
It is noted that the $t$, $\theta$ and $\phi$-components of the 
S force
vanish after summing over $m$ modes. 
Using the formulas for summation over $m$ 
\begin{eqnarray}
\sum_m {2\,\pi \over L} m^2 \,|Y_{\ell m}(\pi/2, \,0 )|^2
&=& {{\cal L}^{(2)} \over 2} \,, \cr 
\sum_m {2\,\pi \over L}|\partial_{\theta}
\,Y_{\ell m}(\pi/2 , \,0 )|^2
&=& {{\cal L}^{(2)} \over 2} \,.
\end{eqnarray}
the $r$-component of the S force is derived as 
\begin{eqnarray}
\left. F^t_{{\rm S,H}}\right|_{\ell}
&=& \left. F^{\theta}_{{\rm S,H}}\right|_{\ell} 
= \left. F^{\phi}_{{\rm S,H}}\right|_{\ell}
= 0 \,, \cr
\left. F^{r(\pm)}_{{\rm S,H}}\right|_{\ell}
&=& 
\mp {\displaystyle \frac {1}{2}}
\,{\displaystyle \frac {\mu ^{2}\,(2\,r_0 - 3\,M)}
{r_0^{3}}} \,L
- {\displaystyle \frac {1}{8}}
\,{\displaystyle \frac {\mu ^{2}\,(4\,r_0 - 7\,M)}{r_0^{3}}}
\nonumber \\ && 
+{\mu^2 \,M \,(172\,L^4-14784\,L^2+299)
\over 128\,r_0^3 (L^2-1)(L^2-4)(L^2-9)}
\nonumber \\ 
&=& \mp {\displaystyle \frac {1}{2}}
\,{\displaystyle \frac {\mu ^{2}\,(2\,r_0 - 3\,M)}
{r_0^{3}}} \,L
- {\displaystyle \frac {1}{8}}
\,{\displaystyle \frac {\mu ^{2}\,(4\,r_0 - 7\,M)}{r_0^{3}}}
+O({1 \over L^2})
\,.
\end{eqnarray}
This is indeed of the standard form.
It should be noted that the factor ${\cal L}^{(2)}$ which is
present in the denominators before summing over $m$ is
cancelled by the same factor that arises from
summation over $m$. If it were present in the final result,
we would not be able to conclude that the summation of
$D^\mu_\ell$ over $\ell$ vanishes.
We note that, apart from the fact that the 
denominator of the $D^\mu_\ell$ term takes the standard form,
the numerical coefficients appearing in the numerator should not be
taken rigorously. This is because our calculation is accurate only
to $O(y^0)$ of the S force, while the numerical coefficients depend
on the $O(y)$ behavior of it.
It is also noted that the $O(1/L)$-terms are absent
in the S force, implying the absence of logarithmic divergence.


\subsection{The S part in the Regge-Wheeler gauge} 
\label{sec:trans}

Now, we transform the S part of the metric perturbation
from the harmonic gauge to the RW gauge.
The gauge transformation functions are
given in the tensor-harmonic expansion form as
\begin{eqnarray}
\xi_{\mu}^{{\rm (odd)}} &=& \sum_{\ell m}
\Lambda_{\ell m}^{{\rm S,H} \to {\rm RW}}(t,\,r)
\left\{ 0, 0, \frac{-1}{\sin\theta}\partial_{\phi}Y_{\ell m}(\theta,\phi),
\sin\theta\partial_{\theta}Y_{\ell m}(\theta,\phi) \right\} \,,
\cr
\xi_{\mu}^{{\rm (even)}} &=& \sum_{\ell m}
\biggl\{M_{0\ell m}^{{\rm S,H} \to {\rm RW}}(t,\,r)
Y_{\ell m}(\theta,\phi),
M_{1\ell m}^{{\rm S,H} \to {\rm RW}}(t,\,r)Y_{\ell m}(\theta,\phi),
\nonumber \\ && \quad 
M_{2\ell m}^{{\rm S,H} \to {\rm RW}}(t,\,r)
\partial_{\theta}Y_{\ell m}(\theta,\phi),
M_{2\ell m}^{{\rm S,H} \to {\rm RW}}(t,\,r)
\partial_{\phi}Y_{\ell m}(\theta,\phi)
 \biggr\}
 \,.
\label{eq:Gtransgen}
\end{eqnarray}
There are one degree of gauge freedom for the odd part and
three for the even part.
To satisfy the RW gauge condition~(\ref{eq:RWcond}),
we obtain the equations for the gauge functions
that are found to be rather simple, and then we find
\begin{eqnarray}
\Lambda_{\ell m}^{{\rm S,H} \to {\rm RW}}(t,\,r)
&=&{i\over2}h_{2\ell m}^{{\rm S,H}}(t,\,r)\,,
\label{eq:GGT1} \\ 
\cr
M_{2\ell m}^{{\rm S,H} \to {\rm RW}}(t,\,r)
&=&-{r^2\over2}G_{\ell m}^{{\rm S,H}}(t,\,r)\,,
\cr
M_{0\ell m}^{{\rm S,H} \to {\rm RW}}(t,\,r)
&=&-h_{0\ell m}^{{\rm (e)S,H}}(t,\,r)
- \partial_t  M_{2\ell m}^{{\rm S,H} \to {\rm RW}}(t,\,r)\,,
\cr
M_{1\ell m}^{{\rm S,H} \to {\rm RW}}(t,\,r)
&=&-h_{1\ell m}^{{\rm (e)S,H}}(t,\,r)
- r^2 \,\partial_r
\left({M_{2\ell m}^{{\rm S,H} \to {\rm RW}}(t,\,r)
\over r^2}\right)\,.
\label{eq:GGT4}
\end{eqnarray}
We note that it is not necessary to calculate any integration with respect to 
$t$ or $r$. It is also noted that the gauge functions are determined uniquely.
This is because the RW gauge is a gauge in which there is no
residual gauge freedom (for $\ell\geq2$).

Then, the S part of the metric perturbation in the RW gauge
is expressed in terms of those in the harmonic gauge as
follows. The odd parity components are found as
\begin{eqnarray}
h_{0\ell m}^{\rm S,RW}(t,\,r)
&=& h_{0\ell m}^{\rm S,H}(t,\,r)
+ \partial_t \,\Lambda_{\ell m}^{{\rm S,H} \to {\rm RW}}(t,\,r) \,,
\nonumber\\
h_{1\ell m}^{\rm S,RW}(t,\,r)
&=& h_{1\ell m}^{\rm S,H}(t,\,r)
+ r^2 \,\partial_r \,
\left(
{\Lambda_{\ell m}^{{\rm S,H} \to {\rm RW}}(t,\,r)
\over r^2}\right) \,,
\label{eq:GTodd}
\end{eqnarray}
and the even parity components are found as
\begin{eqnarray}
H_{0\ell m}^{\rm S,RW}(t,\,r)
&=& H_{0\ell m}^{\rm S,H}(t,\,r)
\nonumber \\ && \hspace{-30mm} 
+{2\,r \over r-2\,M}
\left[ \partial_t
\,M_{0\ell m}^{{\rm S,H} \to {\rm RW}}(t,\,r)
-{M(r-2\,M) \over r^3}
M_{1\ell m}^{{\rm S,H} \to {\rm RW}}(t,\,r)\right] \,,
\nonumber\\
H_{1\ell m}^{\rm S,RW}(t,\,r)
&=& H_{1\ell m}^{\rm S,H}(t,\,r)
\nonumber \\ && \hspace{-30mm} 
+ \left[ \partial_t
\,M_{1\ell m}^{{\rm S,H} \to {\rm RW}}(t,\,r)
+\partial_r \,M_{0\ell m}^{{\rm S,H} \to {\rm RW}}(t,\,r)
-{2\,M \over r(r-2\,M) }
M_{0\ell m}^{{\rm S,H} \to {\rm RW}}(t,\,r)\right] \,,
\nonumber\\
H_{2\ell m}^{\rm S,RW}(t,\,r)
&=& H_{2\ell m}^{\rm S,H}(t,\,r)
\nonumber \\ && \hspace{-30mm} 
+{2(r-2\,M) \over r}
\left[ \partial_r
\,M_{1\ell m}^{{\rm S,H} \to {\rm RW}}(t,\,r)
+{M \over r(r-2\,M)}
M_{1\ell m}^{{\rm S,H} \to {\rm RW}}(t,\,r)\right] \,,
\nonumber\\
K_{\ell m}^{\rm S,RW}(t,\,r)
&=& K_{\ell m}^{\rm S,H}(t,\,r) +{2(r-2\,M) \over r^2}
M_{1\ell m}^{{\rm S,H} \to {\rm RW}}(t,\,r)
\,,
\label{eq:GTeven}
\end{eqnarray}
where the gauge functions are given by Eqs.~(\ref{eq:GGT1})
 and (\ref{eq:GGT4}).

Inserting the S part of the metric perturbation obtained
as described in the previous subsection
to Eqs.~(\ref{eq:GGT1}) and (\ref{eq:GGT4}),
we obtain the gauge functions that transform the S part from the
harmonic gauge to the RW gauge.  
It may be noted that the gauge functions do not
contribute to the metric at the Newtonian order.
In other words, both the RW gauge and the harmonic gauge reduce
to the same (Newtonian) gauge in the Newtonian limit.
The S part of the metric perturbation in the RW gauge is now
found in the form, for example, 
\begin{eqnarray}
h_{0\ell m}^{\rm S,RW}(t,\,r)
&=& {2 \over L}\,\pi\,\mu \biggl[
{\displaystyle \frac 
{4\,i\,T \,m\,r_0\,(L^{2} - 2)\,(u^{\phi})^2}
{{\cal L}^{(2)}\,(L^2-1)}}  
+\cdots\biggr] 
\partial_{\theta}\,Y_{\ell m}^*(\theta_0,\,\phi_0) \,. 
\label{eq:rws6}
\end{eqnarray}

Next we calculate the S part of the self-force.
Of course, it diverges in the coincidence limit. However,
as we noted several times, in the mode-decomposition regularization 
in which the regularization is done for each harmonic mode ($\ell$-mode),
the harmonic coefficients of the S part are finite. 
The calculation is straightforward. We find that the $t$, $\theta$ and 
$\phi$-components of the S force vanish after summing over $m$. 
Summing the above over $m$, the $r$-component of the S force 
is derived as
\begin{eqnarray}
\left. F^t_{{\rm S,RW}}\right|_{\ell} 
&=& 
\left. F^{\theta}_{{\rm S,RW}}\right|_{\ell} 
= \left. F^{\phi}_{{\rm S,RW}}\right|_{\ell} 
= 0 \,, \cr
\left. F^{r(\pm)}_{{\rm S,RW}}\right|_{\ell}
&=& \left. F^{r(\pm)}_{{\rm S,H}}\right|_{\ell} \,.
\end{eqnarray}
We now see that the S force in the RW gauge also has the standard
form as in the case of the harmonic gauge and there is no
$O(1/L)$ term. This is because the difference of the S force between 
the harmonic and RW gauge arises from 2PN order as 
\begin{eqnarray}
\delta F^{r}_{{\rm S,H} \to {\rm RW}}|_\ell &=& \sum_m \, \Biggl[ 
- 3\,{\displaystyle \frac {M\,(r_0 - 2\,M)^{2}\,
h_{1 \ell m}^{\rm (e) S,H}(t_0,r_0)} {r_0^{4}\,(r_0 - 3\,M)}} 
\nonumber \\ &&
+ {\displaystyle \frac {3}{2}} 
\,{\displaystyle \frac {M\,(r_0 - 2\,M)^{2}\,\,
\partial_r G_{\ell m}^{\rm S,H}(t_0,r_0)} 
{r_0^{2}\,(r_0 - 3\,M)}} \Biggr] \,Y_{\ell m}(\theta_0,\,\phi_0)\,. 
\end{eqnarray}

\subsection{Regularized gravitational self-force to 1PN order}
 \label{sec:result}

In the previous two subsections,
we calculated the full and S parts of the self-force
in the RW gauge. Now we are ready to evaluate the regularized
self-force. But there is one more issue to be discussed,
namely, the treatment of the $\ell=0$ and $1$ modes.

The full metric perturbation and its self-force are
derived by the Regge-Wheeler-Zerilli formalism. This means
they contain only the harmonic modes with $\ell \geq 2$.
If we could know the exact S part, then the knowledge of
the modes $\ell\geq2$ would be sufficient to derive the
regular, R part of the self-force, because the R part of
the metric perturbation is known to satisfy the homogeneous
Einstein equations~\cite{Detweiler:2002mi}, and
because there are no non-trivial homogeneous solutions
in the $\ell=0$ and $1$ modes. To be more precise, apart of
the gauge modes that are always present,
the $\ell=0$ homogeneous solution corresponds
to a shift of the black hole mass and the $\ell=1$ odd parity to
adding a small angular momentum to the black hole,
both of which should be put to zero in the absence of 
an orbiting particle. As for the $\ell=1$ even mode,
it is a pure gauge that corresponds to a dipolar shift
of the coordinates.
In other words, apart from possible gauge mode contributions,
the $\ell=0$ and $1$ modes of the full force should be exactly
cancelled by those of the S part.
In reality, however, what we have in hand is only an approximate
S part. In particular, its individual harmonic coefficients do not
have physical meaning. Let us denote the harmonic coefficients of
the approximate S force by $F_\ell^{\rm S,Ap}$, while
the exact S force and the full force by $F_\ell^{\rm S}$
and $F_\ell^{\rm full}$, respectively.
Then, the R force $F^R$ may be expressed as
\begin{eqnarray}
F^{R} &=& \sum_{\ell \geq 2}
\left(F^{\rm full}_{\ell} -F^{\rm S}_{\ell}\right)
=
\sum_{\ell \geq 0}
\left(F^{\rm full}_{\ell} -F^{\rm S}_{\ell}\right)
\nonumber \\
&=&
\sum_{\ell \geq 0}
\left(F^{\rm full}_{\ell} -F^{\rm S,Ap}_{\ell}\right)
-\sum_{\ell\geq0} D_{\ell}
\nonumber \\
&=&
\sum_{\ell \geq 2}
\left(F^{\rm full}_{\ell} -F^{\rm S,Ap}_{\ell}\right)
+\sum_{\ell =0,1}
\left(F^{\rm full}_{\ell} -F^{\rm S,Ap}_{\ell}\right)
\,,
\label{eq:Fldecomp}
\end{eqnarray}
where $D_\ell=F_\ell^{\rm S}-F_\ell^{\rm S,Ap}$, and
the last line follows from the fact that
$F_\ell^{\rm S,Ap}$ are assumed to be obtained from a
sufficiently accurate spherical extension of the local
behavior of the S part to guarantee $\sum_{\ell\geq0}D_\ell=0$.
Thus, it is necessary to evaluate the $\ell=0$ and $1$ modes
of the full force to evaluate the self-force correctly.

First, we consider the contributions of $\ell \geq 2$ to
the self-force. As noted before, for the 1PN calculation,
the only $r$-component of the full and S part of the self-force
is non-zero.
The $\ell$ mode coefficients
corresponding to the first term in the last line of
Eq.~(\ref{eq:Fldecomp}) are derived as
\begin{eqnarray}
\left. F^{r}_{{\rm RW}}\right|_{\ell} &=&
\left. F^{r}_{{\rm full,RW}}\right|_{\ell}
-\left. F^{r}_{{\rm S,RW}}\right|_{\ell} \nonumber \\ 
&=& -{45\,\mu^2\, M\over 8 (2\ell-1) (2\ell +3) \,r_0^3} \,.
\end{eqnarray}
Summing over $\ell$ modes, we obtain
\begin{eqnarray}
F^r_{{\rm RW}}(\ell \geq 2) = -{3\,\mu^2\, M \over 4 \,r_0^3} \,.
\label{eq:FRW2}
\end{eqnarray}

Next, we consider the $\ell=0$ and $1$ modes. 
It is noted that the $\ell=0$ and $\ell=1$ odd modes, 
which describe the perturbation in the total mass and angular momentum,
respectively, of the system due to the presence of the particle,
are determinable in the harmonic gauge, with the retarded boundary
condition. On the other hand, we were unable to solve for 
the $\ell=1$ even mode in the harmonic gauge.
Since it is locally a gauge mode describing a shift of the center
of mass coordinates, this gives rise to an ambiguity in the
final result of the self-force. Nevertheless,
we were able to resolve this ambiguity at Newtonian order,
and hence to obtain an unambiguous interpretation of
the resulting self-force.

The correction to the regularized self-force that
arises from the $\ell=0$ and $\ell=1$ modes, 
corresponding to the second term in the last line of
Eq.~(\ref{eq:Fldecomp}), is found as~\cite{Nakano:2003he,DP}  
\begin{eqnarray}
\delta F^r_{{\rm RW}}(\ell =0,\,1) = {2\,\mu^2 \over r_0^2}
-{57\,\mu^2 M \over 4 r_0^3} \,.
\label{eq:FRW01}
\end{eqnarray}

Finally, adding Eqs.~(\ref{eq:FRW2}) and (\ref{eq:FRW01}),
we obtain the regularized gravitational self-force to
the 1PN order as
\begin{eqnarray}
F^r_{{\rm RW}} = {2\,\mu^2 \over r_0^2}
-{15\,\mu^2 M \over r_0^3} \,.
\end{eqnarray}
Since there will be no effect of the gravitational radiation at the
1PN order, i.e., the $t$- and $\phi$-components are zero,
the above force describes the correction to the radius of the orbit
that deviates from the geodesic on the unperturbed background.
It is noted that the first term proportional to $\mu^2$ is just
the correction to the total mass of the system at the Newtonian
order, where $r_0$ is interpreted as the distance from the center of
mass of the system to the particle.
Although this is a bit cumbersome way to obtain the Newtonian
force, it is nevertheless an important result because it 
was derived from the relativistic self-force for the first time.

\section{New Regularization}\label{seq:newreg}

As mentioned in the previous section, the
frequency spectrum becomes monochromatic in the circular case
so that the frequency integral can be trivially performed
to give the harmonic coefficients of the full metric
perturbation in the time-domain.
But when we consider the general orbit, the frequency spectrum
is highly non-trivial and the frequency integral 
becomes formidable to perform analytically,
making the S part subtraction difficult to carry out.
In this section, we present a new analytical method
that can circumvent this difficulty~\cite{HJNSST}.

Here, again for simplicity, we go back to
the case of a particle with a scalar charge $q$,
and consider the self-force given by
$F_{\alpha}[\varphi]$ defined in Eq.~(\ref{eq:force}).
We first describe a new decomposition, which we call
the $\tilde{\rm S}$-$\tilde{\rm R}$ decomposition, of the
Green function in the Fourier-harmonic domain. Next
we show that only the $\tilde{\rm S}$ part needs to be
regularized, but it can be easily transformed to an expression
in the time domain if we employ the post Newtonian expansion.
We then calculate the ($\tilde{\rm S}-{\rm S}$) part,
which is finite and regular, for general orbit to 6PN order
and for circular orbit to 18PN order up to now,
although we present the explicit result only to 4PN order
in this paper. Finally, specializing to the case of circular
orbits, for which we can calculate the $\tilde{\rm R}$ part
analytically as well, we evaluate the self-force for
several radii. In particular, we compare our result with
the result numerically obtained by
Detweiler, Messaritaki, Whiting and 
Diaz-Rivera~\cite{Detweiler:2002gi,Diaz-Rivera:2004ik}.

\subsection{The $\tilde{S}$-$\tilde{R}$ decomposition}
The full scalar field induced by a scalar-charged particle is given
by Eq.~(\ref{eq:full-g}) in terms of the retarded Green function
$G^{\rm full}(x,x')$. The retarded Green function is
represented in terms of the Fourier-harmonic decomposition as
\begin{eqnarray}
G^{\rm full}(x,x') &=& \int {d\omega\over2\pi} \,
e^{-i\omega(t-t')} \sum_{\ell m} g^{\rm full}_{\ell m\omega}(r,r')
Y_{\ell m}(\theta,\phi)Y^*_{\ell m} (\theta',\phi')\,.
\label{eq:FHdeco-full-g}
\end{eqnarray}
Then, the Klein-Gordon equation~(\ref{eq:wave}) reduces to
 an ordinary differential equation for the radial Green function as
\begin{eqnarray}
&&\left[\left(1-{2M\over r}\right){d^2\over dr^2} +{2(r-M)\over
r^2}{d\over dr}+\left(\frac{r\omega^2}{r-2M}
-{\ell(\ell+1)\over r^2}\right)
\right]
g^{\rm full}_{\ell m\omega}(r,r')\cr
&&\hspace*{5cm}
 = -{1\over r^2}\delta(r-r')   \,.
 \label{eq:radial-eq}
\end{eqnarray}

The radial part of the full Green function can be expressed in
terms of homogeneous solutions of Eq.~(\ref{eq:radial-eq}), which
can be obtained using a systematic analytic method developed in
Ref.~\cite{Mano2}. We have
\begin{eqnarray}
g_{\ell m\omega}^{\rm full}(r,r')
={-1\over W_{\ell m\omega}(\varphi_{\rm in}^{\nu},\varphi_{\rm up}^{\nu})}
\nonumber\\
\qquad\times 
 \left(\varphi_{\rm in}^{\nu}(r)\varphi_{\rm up}^{\nu}(r')\theta(r'-r) 
+\varphi_{\rm up}^{\nu}(r)\varphi_{\rm in}^{\nu}(r')\theta(r-r')\right) \,;
\nonumber\\
W_{\ell m\omega}(\varphi_{\rm in}^{\nu},\varphi_{\rm up}^{\nu})
\nonumber\\
\qquad=r^2\left(1-{2M\over r}\right) 
\left[\biggl({d\over dr}\varphi_{\rm up}^{\nu}(r)\biggr)
\varphi_{\rm in}^{\nu}(r) -\biggl({d\over dr}\varphi_{\rm in }^{\nu}(r)\biggr)
\varphi_{\rm up}^{\nu}(r) \right].
\end{eqnarray}
Here, the in-going and up-going homogeneous solutions are
denoted, respectively, by $\varphi_{\rm in}^{\nu}$ and 
$\varphi_{\rm up}^{\nu}$, and $\nu$ is called the `renormalized angular
momentum'~\cite{Mano2,Mano:1996vt,ManoTak}, which is equal to $\ell$ in
the limit $M\omega \to0$.

We express the homogeneous solutions $\varphi_{\rm in}^\nu$ and
$\varphi_{\rm up}^\nu$ in terms of the Coulomb wave functions 
$\varphi_{\rm c}^{\nu}$ and $\varphi_{\rm c}^{-\nu-1}$
~\cite{Mano:1996vt,ManoTak,Sasaki-Tagoshi};
\begin{eqnarray}
\varphi_{\rm in}^{\nu} &=& \varphi_{\rm c}^{\nu} + \tilde{\beta}_{\nu}\,
\varphi_{\rm c}^{-\nu-1} \,,\quad
\varphi_{\rm up}^{\nu} = \tilde{\gamma}_{\nu}\,\varphi_{\rm c}^{\nu}
+\varphi_{\rm c}^{-\nu-1}.
\end{eqnarray}
The properties and the relations of the coefficients
$\tilde{\beta}_\nu,\,\tilde{\gamma}_\nu$ are studied in
Ref.~\cite{Mano:1996vt} and~\cite{ManoTak}. An essential point 
to be noted here is that when we consider the post Newtonian
expansion, i.e., when $\varphi_c^\nu$ is expanded in terms of $z:=\omega r$
and $\epsilon:=2M\omega$, 
$\Phi^\nu:=(2z)^{-\nu} \varphi_c^{\nu}$ contains only terms that are
integer powers of $z$ and $\epsilon$.
Furthermore, this PN expansion turns out to be a double Taylor
series expansion in $z^2$ and $\epsilon/z$, i.e., there appear
only positive powers of $\omega^2$. 

We now divide the Green function into two parts, as
\begin{eqnarray}\label{split}
&&g_{\ell m\omega}^{\rm full}(r,r')
= g_{\ell m\omega}^{\tilde{\rm S}}(r,r')
 +g_{\ell m\omega}^{\tilde{\rm R}}(r,r') \,,
\end{eqnarray}
where
\begin{eqnarray}
g_{\ell m\omega}^{\tilde{\rm S}}(r,r')
&=&
 {-1\over W_{\ell m\omega}(\varphi_{\rm c}^{\nu},\varphi_{\rm c}^{-\nu-1})}
\cr
&&
\times \biggl[
 \varphi_{\rm c}^{\nu}(r)\varphi_{\rm c}^{-\nu-1}(r')\theta(r'-r)
+\varphi_{\rm c}^{-\nu-1}(r)\varphi_{\rm c}^{\nu}(r')\theta(r-r') \biggr],
\label{eq:gC} \\
g_{\ell m\omega}^{\tilde{\rm R}}(r,r')&=& {-1\over
(1-\tilde\beta_\nu\tilde\gamma_\nu) W_{\ell m\omega}
  (\varphi_{\rm c}^{\nu},\varphi_{\rm c}^{-\nu-1})}
\cr&&
\times \biggl[\tilde\beta_\nu\tilde\gamma_\nu
 \left(\varphi_{\rm c}^{\nu}(r)\varphi_{\rm c}^{-\nu-1}(r')
 +\varphi_{\rm c}^{-\nu-1}(r)\varphi_{\rm c}^{\nu}(r')\right) 
\cr
&&
+\tilde\gamma_\nu \varphi_{\rm c}^{\nu}(r)\varphi_{\rm c}^{\nu}(r')
 +\tilde\beta_\nu \varphi_{\rm c}^{-\nu-1}(r)\varphi_{\rm c}^{-\nu-1}(r')
 \biggr]\,. \label{eq:gR}
\end{eqnarray}
Using results derived in Ref.~\cite{Mano:1996vt}, we obtain the behavior of
the coefficients $\{\tilde\beta_\nu,\,\tilde\gamma_\nu\}$ in the PN
expansion as
\begin{eqnarray}
\tilde\beta_\nu=O(v^{6\ell+3})\,, \quad
\tilde\gamma_\nu=O(v^{-3})\,, \label{eq:betagamma}
\end{eqnarray}
where we have set $z=O(v)$ and $\epsilon=O(v^3)$
with $v$ being the characteristic orbital velocity.
The functions $\varphi^\nu_c$ and $\varphi^{-\nu-1}_c$ are,
respectively, of $O(v^\ell)$ and $O(v^{-\ell-1})$ (except at
$\ell=0$).
Therefore, the three terms in the $\tilde{\rm R}$ part of
the Green function become, respectively, of $O(v^{6\ell})$,
$O(v^{2\ell-2})$ and $O(v^{4\ell+2})$ relative to the
 $\tilde{\rm S}$ part.

As should be clear from Eqs.~(\ref{eq:gC}) and (\ref{eq:gR}),
the $\tilde{\rm R}$ part is regular at $r=r'$, hence the part that 
needs to be regularized is the $\tilde{\rm S}$ part.
Furthermore, if we truncate the PN expansion at a finite order,
the $\tilde{\rm R}$ part terminates at finite $\ell$.
Therefore the expression for the R part of the self-force in new
decomposition takes following form, 
\begin{eqnarray}
 F^{\rm R}_{\alpha}
&=& F^{\rm full}_{\alpha}-F^{\rm S}_{\alpha}
\nonumber \\ 
&=& (F^{\tilde{\rm S}}_{\alpha}
-F^{\rm S}_{\alpha})
+\sum_{\ell=0}^{\ell_{\rm{max}}}\left. F^{\tilde{\rm R}}_{\alpha}\right|_{\ell} \,.
\label{eq:Rforceform}
\end{eqnarray}

\subsection{Computation of the $\tilde{S}$ part}
\label{sec:tilde-S}

We now compute the force due to the $\tilde{\rm S}$ part for a general
orbit.
The difference between the S force and the $\tilde{\rm S}$ force
should be finite, because the $\tilde{\rm R}$ force
is finite.
Thus, in general, 
when expanded in terms of the spherical harmonics, 
the $\tilde{\rm S}$ force must take the same form as the
S force
\begin{equation}
\left. F^{\tilde{\rm S}}_{\alpha}\right|_{\ell}
  =A_\alpha L+B_\alpha + \tilde D_{\alpha,\ell} \,.
\label{standardform}
\end{equation}
Below we confirm explicitly that both $A_{\alpha}$ and $B_{\alpha}$ for
the $\tilde{\rm S}$ force coincide with those of
Eqs.~(\ref{eq:Aterm}) and (\ref{eq:Bterm}).
 Therefore, the $\tilde{\rm S}$ force
minus the S force, which is finite, is given by
\begin{equation}
F^{\tilde{\rm S}-{\rm S}}_{\alpha}\equiv
 \sum_{\ell=0}^{\infty} 
\left(\left. F^{\tilde{\rm S}}_{\alpha}\right|_{\ell}
   -\left. F^{\rm S}_{\alpha}\right|_{\ell}\right)
   = \sum_{\ell=0}^{\infty} \tilde D_{\alpha,\ell}.
\end{equation}

To obtain an expression for the $\tilde{\rm S}$ force in the form
of Eq.~(\ref{standardform}), it is necessary to perform the $\omega$
integration explicitly. Here, the key fact is that there appears no
fractional power of $\omega$ in the $\tilde{\rm S}$ part. This is
because we have chosen $\phi_c^\nu$ and $\phi_c^{-\nu-1}$ as the
two independent basis functions. As noted above, except for the
overall fractional powers $z^\nu$ and $z^{-\nu-1}$, they contain
only the terms with positive integer powers of $\omega^2$. When we
consider a product of these two functions, $\omega$ contained in
the overall factors $z^\nu$ and $z^{-\nu-1}$ just produces
$\omega^{-1}$, which is canceled by $\omega$ from the inverse of
the Wronskian. Thus, $g^{\tilde{\rm S}}_{\ell m\omega}(r,r')$ is
expanded as
\begin{equation}
 g^{\tilde{\rm S}}_{\ell m\omega}(r,r')
   =\sum_{k=0}^{\infty} \omega^{2k} {\cal G}_{\ell m k}(r,r').
\end{equation}
The Fourier transform of
$\omega^{2n}$ simply produces
\[
 \int d\omega\, \omega^{2n} e^{-i\omega (t-t')}=
    2\pi{ (-1)^n \partial_{t'}^{2n} }\delta (t-t').
\]
Differentiation of the delta function in the expression above can
be integrated by parts to act on the source term. 
Thus, the integration over $\omega$ can be performed easily,
and we can express the $\tilde{\rm S}$ force in the time domain as
\begin{eqnarray}
\left. F_{\alpha}^{\tilde{\rm S}}\right|_{\ell} = q^2 
\lim_{x\to z(t)} P_\alpha{}^\beta\nabla_\beta 
\cr
\qquad
\sum_{m,\,k}  \left[ 
(-1)^k  (\partial_t)^{2k} {d\tau(t)\over dt}
     {\cal G}_{\ell m k}(r,z^r(t)) Y_{\ell m}(\theta, \phi)
     Y_{\ell m}^* (z^\theta (t), z^\phi(t))
\right]\,.
\label{forceS-S}
\end{eqnarray}

The transformation of the $\tilde{\rm S}$ part into the time domain makes it
possible to subtract the divergent S part analytically. 
If we were to perform this
subtraction numerically, the fraction to be subtracted would become
closer and closer to unity as $\ell$ increases. Apparently, this would
mean a stringent requirement on numerical accuracy.
In this sense, we anticipate a clear advantage in the analytical
subtraction.

\subsection{The $(\tilde{S}-{S})$ force}
\label{sub:S-tildeS}

The result for the $\tilde{S}$ force is
\begin{eqnarray}
\label{eq:ell-mode-result}
 && \left. F_{t}^{\tilde{\rm S}}\right|_{\ell}=\frac{q^2 u^r}{4\pi r_0^2}
\sum_{n=0}^{\infty}K_{t,\,\ell}^{(n)} \,,\quad
 \left. F_{\theta}^{\tilde{\rm S}}\right|_{\ell}=0,\quad
 \left. F_{\phi}^{\tilde{\rm S}} \right|_{\ell}=\frac{q^2 u^r\mathcal{L}}{4\pi r_0^2}
 \sum_{n=0}^{\infty}K_{\phi,\,\ell}^{(n)} \,,
\end{eqnarray}
and
\begin{equation}
\left. F_{r}^{\tilde{\rm S}}\right|_{\ell}
=-\frac{\mathcal{E}}{u^r(1-2M/r_0)}
\left. F_{t}^{\tilde{\rm S}}\right|_{\ell}
-\frac{\mathcal{L}}{u^r r_0^2}
\left. F_{\phi}^{\tilde{\rm S}}\right|_{\ell} \,,
\end{equation}
where the coefficients $K_{\alpha,\,\ell}^{(n)}$, whose superscript 
$(n)$ represents the PN order, are formally given by
\begin{eqnarray}
&& K_{t,\,\ell}^{(n)}=\sum_{i+j+k=n} d_{t,\,\ell}^{\ (ijk)}\
(\delta_{\mathcal{E}})^i
\left(\frac{\mathcal{L}^{2}}{r_0^2}\right)^jU^k \,,\cr
&& K_{\phi,\,\ell}^{(n)}=\sum_{i+j+k=n}
d_{\phi,\,\ell}^{\ (ijk)}\  (\delta_{\mathcal{E}})^i
\left(\frac{\mathcal{L}^{2}}{r_0^2}\right)^jU^k \,.
\end{eqnarray}
Here, the quantities $d_{\alpha,\,\ell}^{\ (ijk)}$ are some functions of 
$\ell$, $r_0\equiv z^r(t_0)$,
$\delta_{\mathcal{E}}\equiv1-\displaystyle\frac{1}{\mathcal{E}^2}$,
$U\equiv\displaystyle\frac{M}{r_0},$ and $\mathcal{E}$ and
$\mathcal{L}$ are, respectively, the energy and the angular
momentum of the particle. To obtain these expressions, we have
used the first integrals of the geodesic equations
and we have also reduced higher-order derivatives with respect to $t$ by
using the equations of motion.

To summarize, the $(\tilde{\rm S}-{\rm S})$ force is given by
\begin{eqnarray}
 F^{\tilde{S}-S}_\alpha &=& \sum_{\ell=0}^{\infty}
\left(\left. F^{\tilde{S}}_{\alpha}\right|_{\ell}-\left. F^{S}_{\alpha}\right|_{\ell}\right)
\nonumber \\ 
&=&  \sum_{\ell=0}^{\infty}
\left(\left. F^{\tilde{S}}_{\alpha}\right|_{\ell}-
A_{\alpha}\left(\ell+\frac{1}{2}\right)-B_{\alpha}\right).
\end{eqnarray}
The final result is
\begin{eqnarray}
\label{eq:tilS-S-result}
 && F_t^{\tilde{\rm S}-{\rm S}}=\frac{q^2 u^r}{4\pi r_0^2}
\sum_{n=0}^{\infty}C_{t}^{\tilde{\rm S}-{\rm S}(n)},\quad
 F_\theta^{\tilde{\rm S}-{\rm S}}=0,\quad
 F_\phi^{\tilde{\rm S}-{\rm S}}
=\frac{q^2 u^r\mathcal{L}}{4\pi r_0^2}
 \sum_{n=0}^{\infty}C_\phi^{\tilde{\rm S}-{\rm S}(n)},\quad
\end{eqnarray}
and
\begin{equation}
F_r^{\tilde{\rm S}-{\rm S}}
=-\frac{\mathcal{E}}{u^r(1-2M/r_0)}F_t^{\tilde{\rm S}-{\rm S}}
-\frac{\mathcal{L}}{u^r r_0^2}F_\phi^{\tilde{\rm S}-{\rm S}},
\end{equation}
where the coefficients $C_\alpha^{\tilde{\rm S}-{\rm S}(n)}$,
 whose upper index, $n$, represents the PN order.
The calculation of $C_\alpha^{\tilde{\rm S}-{\rm S}(n)}$ to
a given order of post Newtonian expansion is straightforward.
Most importantly, once we compute them, we can just keep
the results and apply them to any general orbit at any time.
Up to now, we have computed the ($\tilde{\rm S}-{\rm S}$) force
to 6 PN order. As example, we list some low PN orders terms:
\begin{eqnarray*}
 C^{\tilde{\rm S}-{\rm S}(0)}_t&=&\frac{73}{133},\cr
 C^{\tilde{\rm S}-{\rm S}(1)}_t&=&-\frac{610}{31521}\delta_{\mathcal{E}}
+\frac{282}{1501}\frac{\mathcal{L}^2}{r_0^2}
-\frac{59590}{31521}U,\cr
 C^{\tilde{\rm S}-{\rm S}(2)}_t
&=&-\frac{2296958}{8878415}\delta_{\mathcal{E}}^2
+\left[\frac{14127898}{8878415}\frac{\mathcal{L}^2}{r_0^2}
-\frac{20571064}{26635245}U\right]\delta_{\mathcal{E}}\cr&&
-\frac{5579893}{1775683}\frac{\mathcal{L}^4}{r_0^4}
-\frac{59116}{253669}\frac{\mathcal{L}^2U}{r_0^2}
-\frac{18112}{10507}U^2,\cr
 C^{\tilde{\rm S}-{\rm S}(3)}_t
&=&-\frac{115291414894}{269415503175}\delta_{\mathcal{E}}^3
+\left[\frac{43471970326}{17961033545}\frac{\mathcal{L}^2}{r_0^2}
-\frac{48448379368}{89805167725}U\right]\delta_{\mathcal{E}}^2\cr&&
+\biggl[-\frac{22584903396}{2565861935}\frac{\mathcal{L}^4}{r_0^4}
-\frac{508295808}{2565861935}\frac{\mathcal{L}^2U}{r_0^2}
\cr&&
-\left(\frac{244692415685084}{8336485426815}
+\frac{21\pi^2}{32}\right)U^2\biggr]\delta_{\mathcal{E}}
\cr&&
+\frac{16156048}{1301367}\frac{\mathcal{L}^6}{r_0^6}
-\frac{291581166}{32479265}\frac{\mathcal{L}^4U}{r_0^4}\cr&&
+\left(\frac{122513312775814}{1667297085363}+\frac{105\pi^2}{64}\right)
\frac{\mathcal{L}^2U^2}{r_0^2}\cr&&
-\left(\frac{1707952144915294}{25009456280445}+\frac{7\pi^2}{4}\right)U^3,\cr
 \cr
 \cr
C^{\tilde{\rm S}-{\rm S}(0)}_\phi&=&\frac{960}{10507},\cr
C^{\tilde{\rm S}-{\rm S}(1)}_\phi
&=&\frac{23145656}{26635245}\delta_{\mathcal{E}}
+\frac{33594}{1775683}\frac{\mathcal{L}^2}{r_0^2}
+\frac{2403368}{761007}U,\cr
C^{\tilde{\rm S}-{\rm S}(2)}_\phi
&=&\frac{46189522292}{53883100635}\delta_{\mathcal{E}}
+\left[-\frac{94186639}{2565861935}\frac{\mathcal{L}^2}{r_0^2}
+\frac{1182191716}{2565861935}U\right]
\delta_{\mathcal{E}}\cr&&
-\frac{86499760}{39474799}\frac{\mathcal{L}^4}{r_0^4}
+\frac{965295376}{513172387}\frac{\mathcal{L}^2 U}{r_0^2}\cr&&
+\left(-\frac{231720301397372}{8336485426815}-\frac{21\pi^2}{32}\right)U^2.
\end{eqnarray*}

Once we obtain the general expression for the $(\tilde{\rm S}-S)$ part of
the force, computation of the remaining $\tilde{\rm R}$ part is
rather easy, because only terms up to a finite value of $\ell$
contribute to the force for a given PN order.
Then, the R force, which is what we want in the end, is
given by the form (\ref{eq:Rforceform}).

One of the important properties of the $(\tilde{\rm S}-{\rm S})$ force
 is that it contains only the conservative part of the force.
From Eqs.~(\ref{eq:ell-mode-result}), the $\ell$-mode of the
 $\tilde S$ force takes the form
\begin{equation}
\left. F^{\tilde{\rm S}}_{t}\right|_{\ell}={\cal F}_{t,\ell}(r,\mathcal{E},\mathcal{L})\,u^r\,,
\, \left. F^{\tilde{\rm S}}_{r}\right|_{\ell}
={\cal F}_{r,\ell}(r,\mathcal{E},\mathcal{L})\,,
 \, \left. F^{\tilde{\rm S}}_{\phi}\right|_{\ell}
={\cal F}_{\phi,\ell}(r,\mathcal{E},\mathcal{L})\,u^r\,.
\end{equation}
The S part of the force is known to have exactly the same form.
This implies that the $(\tilde{\rm S}-{\rm S})$ force also takes
the same form. Thus, after summing over $\ell$, we conclude that
the final form of the $(\tilde{\rm S}-{\rm S})$ force is
\begin{eqnarray}
F^{\tilde{\rm S}-{\rm S}}_t={\cal F}_t (r,\mathcal{E},\mathcal{L})\,u^r,\ 
F^{\tilde{\rm S}-{\rm S}}_r={\cal F}_r (r,\mathcal{E},\mathcal{L}), \ 
F^{\tilde{\rm S}-{\rm S}}_\phi
={\cal F}_\phi(r,\mathcal{E},\mathcal{L})\, u^r\,.
\label{tilS-Sform}
\end{eqnarray}

We can now explicitly show that the above form of the force
implies the absence of a dissipative reaction effect. In other words,
the force is conservative. The
equations of motion to $O(\mu^2)$ are given by
\begin{equation}
 \mu{D\over d\tau} \tilde u^\mu=F^\mu,
\end{equation}
where $\tilde u^\mu$ is the perturbed four velocity and $D/d\tau$
is the covariant derivative. Then, we obtain the evolution equation for
the perturbed energy $\tilde{\cal E}:=-\mu\,\hat t_{\mu} \tilde u^\mu$ as
\begin{equation}
 {d\tilde{\cal E}\over d\tau}
 =-\mu{D\over d\tau}(\hat t_\mu \tilde u^\mu)
     =-\hat t_\mu F^\mu = -{\cal F}_t (r) {dr\over d\tau},
\end{equation}
where $\hat t^\mu=(\partial_t)^\mu$
is the time-like Killing vector. This equation
is integrated to give
\begin{equation}
  \tilde {\cal E}={\cal E}-\int^r \, {\cal F}_t(r) dr.
\end{equation}
Here, ${\cal E}$ is an integration constant, which we can interpret
as the unperturbed energy.
In the same manner, for the perturbed angular momentum ${\cal L}$, we obtain
\begin{equation}
  \tilde {\cal L}={\cal L}+\int^r \, {\cal F}_\phi (r) dr.
\end{equation}
Thus we find that there is no cumulative effect on the
evolution of the energy and angular momentum of the particle.
In other words,
a force of the form~(\ref{tilS-Sform}) preserves the presence of the
constants of motion ${\cal E}$ and ${\cal L}$.
Concerning the radial motion, $\tilde{u}^r$ can be expressed
in terms of $\mu\,\tilde{u}_t=\tilde{\cal E}$ and
$\mu\,\tilde{u}_\phi=\tilde{\cal L}$
by using the normalization condition of the
four velocity. We have 
\begin{eqnarray}
\mu\,\tilde u^r=\pm\left[
\tilde{\cal E}^2-(1-2M/r)\left(1+\tilde{\cal L}^2/r^2\right)
\right]^{1/2}.
\end{eqnarray}
Thus, $\tilde u^r$ is obtained as a function of $r$.

\subsection{Testing the efficiency for circular orbits}

To examine the efficiency of this new regularization
method, we revisit the problem of the self-force for circular orbits.
For this purpose, we first calculate the $\tilde{\rm S}$ force
for a circular orbit to 18PN order.
Then, combining with the calculation of the $\tilde{\rm R}$ part,
which can be done with sufficient accuracy in order not to
spoil the 18PN order accuracy of the ($\tilde{\rm S}-{\rm S}$) part,
the regularized scalar self-force is evaluated
 and is compared with the result obtained
by Detweiler, Messaritaki and Whiting~\cite{Detweiler:2002gi},
and very recently by Diaz-Rivera et al.~\cite{Diaz-Rivera:2004ik}.

The components of the ($\tilde{\rm S}-{\rm S}$) self-force,
$F_{\alpha}^{\tilde{\rm S}-{\rm S}}
= F_{\alpha}^{\tilde{\rm S}}-F_{\alpha}^{\rm S}$,
have been given for general orbits in
Eq.~(\ref{eq:tilS-S-result}).
Since the ($\tilde{\rm S}-{\rm S}$) force is expressed in terms of 
local quantities of the particle, i.e., its position and velocity,
what we have to do is just to specify the orbit.
In the present case, we consider a circular orbit,
given by Eq.~(\ref{eq:orbit}).
Then we find only the $r$-component is
non-vanishing for the ($\tilde{\rm S}-{\rm S}$) force
 because the $t$- and
$\phi$-components are directly related to rates of change of
the energy and angular momentum, and are purely dissipative
for circular orbits. 
To 4PN order, it is given explicitly by
\begin{eqnarray}
F^{\tilde{\rm S}-{\rm S}}_r
&=& \frac{q^2}{4\pi r_0^2}
\biggl[
-\frac{73}{133}+\frac{16151}{21014}V^2
+\frac{395567}{106808}V^4
\nonumber\\
&&\quad+\left(\frac{1107284037660637}{400151300487120}
 +\frac{7}{64}\pi^2\right)V^6
\nonumber\\
&&\quad+ \left( - \frac
{182118981911377689978271}{8548630707351386171520} + \frac{29\,\pi
^{2}}{1024} \right)\,V^8 \biggr],
\end{eqnarray}
where $V=\sqrt{M/r_0}=r_0 \Omega$. 

The components of the $\tilde{\rm R}$ force are 
formally given by
\begin{eqnarray}
F_t^{\tilde{\rm R}} &=&
-\frac{iq^2\Omega}{u^t}\sum_{\ell m} m \,
g_{\ell m, m\Omega}^{\tilde{\rm R}}(r_0,r_0)
\left|Y_{\ell m}(\frac{\pi}{2},0)\right|^2,
\nonumber \\
F_r^{\tilde{\rm R}} &=&
\frac{q^2}{u^t}\sum_{\ell m} \left.
\partial_r g_{\ell m, m\Omega}^{\tilde{\rm R}}(r,r_0)\right|_{r=r_0}
\left|Y_{\ell m}(\frac{\pi}{2},0)\right|^2,
\nonumber \\
F_\theta^{\tilde{\rm R}} &=& 0,
\nonumber \\
F_\phi^{\tilde{\rm R}} &=&
-\frac{1}{\Omega}F_t^{\tilde{\rm R}}. \label{eq:tilR}
\end{eqnarray}
The $\tilde{\rm R}$ force in this case can be completely
obtained analytically because the integration with
respect to $\omega$ is done by substituting $m \Omega$ for
$\omega$. Also note that if we need the
precision up to $n$-PN order inclusive, it is sufficient for us to
calculate the modes up to $\ell \le n+1$.
 The 4PN results, after summation over $\ell$-modes, are
\begin{eqnarray}
F^{\tilde{\rm R}}_r 
&=& {q^2 \over 4\pi r_0^2} \Biggl[ {73 \over
133}-{16151 \over 21014}V^2-{395567 \over 106808}V^4
\cr 
&&\quad
+\left( -{4
\over 3}\gamma-{4 \over 3}\ln(2V) -{1196206548879997 \over
400151300487120}\right)V^6 
\cr 
&&
\quad+\left(\frac{59372120592232147984979}{1709726141470277234304} -
\frac{14}{3} \,\ln V
\right.\cr
&&\quad\qquad \left.- \frac{66}{5} \,\ln(2) - \frac
{14\,\gamma}{3} \right)V^8\Biggr]\,, 
\cr
F^{\tilde{\rm R}}_t
 &=& {q^2 V\over 4\pi r_0^2} \Biggl[ {1 \over 3}V^3-{1 \over
6}V^5+{2\pi \over 3}V^6 -{77 \over 24}V^7 +{9\pi \over 5}V^8
\Biggr] \,.
\end{eqnarray}
Here $\gamma$ is Euler's constant, $\gamma=0.57\cdots$.
As expected, the $t$-component which represents the energy loss rate
starts at $1.5$PN order, corresponding to dipole radiation.

An important property of the $\tilde{\rm R}$ force is that each
$\ell$-mode of it may be evaluated without performing the
post Newtonian expansion. In other words, we do not
have to expand the $\tilde{\rm R}$ part of the Green function in 
powers of $\omega$. Instead, we can easily compute each $\ell$-mode
with sufficient accuracy for any given radius $r_0$.
Thus, as far as the $\tilde{\rm R}$ part is concerned, what we
actually employ is not the standard post Newtonian expansion,
but the truncation of the series expansion in $\ell$ at
a given $\ell_{\rm max}$ 
compatible with the PN order of our interest.

The total self-force is obtained by summing the
($\tilde{\rm S}-{\rm S}$) force and the $\tilde{\rm R}$ force.
We have computed the force accurate to 18PN order so far.
But here, we present the result explicitly only to 4PN order, 
\begin{eqnarray}
F^{\rm R}_r &=& {q^2 \over 4\pi r_0^2} \biggl[ \left( -{4 \over
3}\gamma+{7 \over 64}\pi^2-{4 \over 3}\ln(2V) -{2 \over
9}\right)V^6 \cr && +\left( {604 \over 45} + {29\pi^2 \over 1024}
- {66 \over 5}\ln(2) - {14 \over 3}\ln V - {14 \over 3}\gamma
\right)V^8
\biggr]  \,,
 \label{eq:FRr_cir}\\
F^{\rm R}_t &=& F^{\tilde{\rm R}}_t.
\end{eqnarray}
In the $r$-component of the scalar self-force there is a
significant cancellation between the $(\tilde{\rm S}-{\rm S})$ part
 and the $\tilde{\rm R}$ part, and the total force begins at 3PN order.

\begin{figure}[ht]
\centering
\epsfxsize=300pt
 \epsfbox{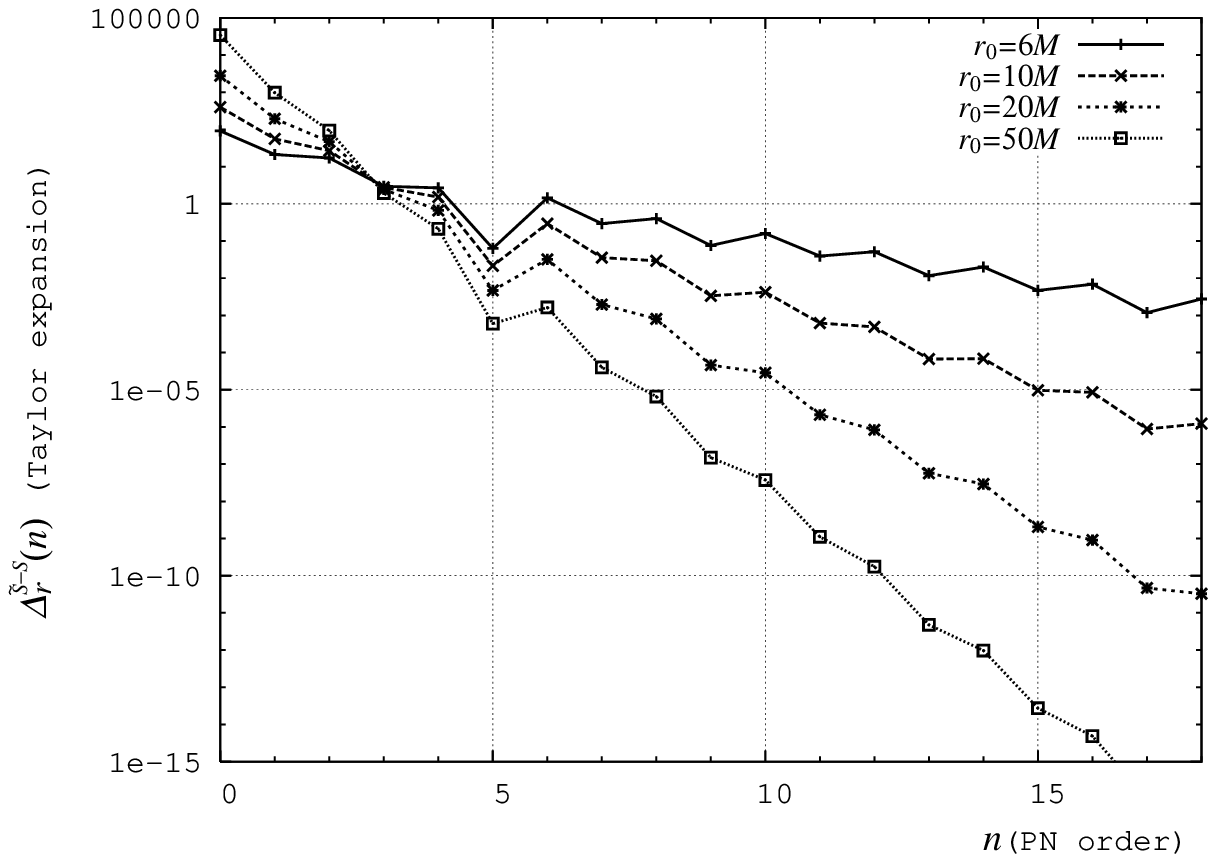}
\epsfxsize=300pt 
 \epsfbox{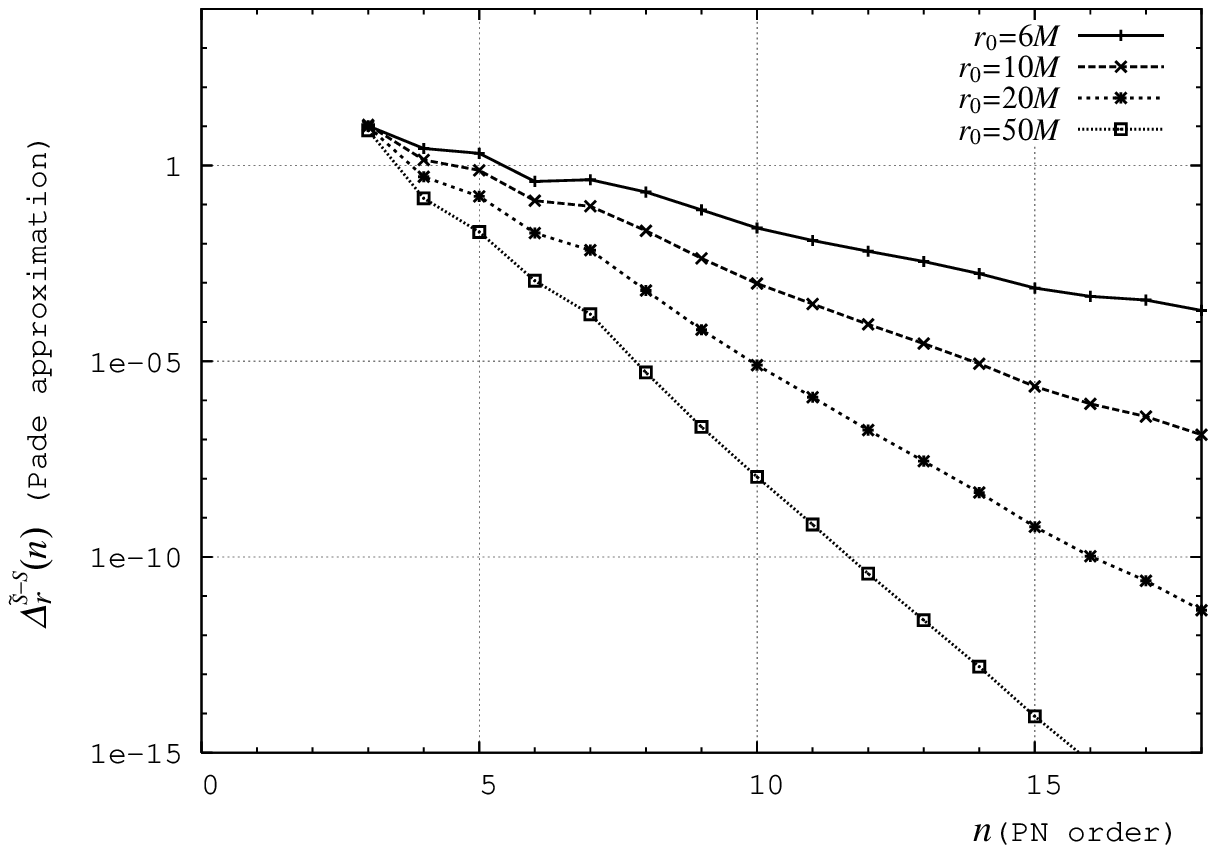}
\caption{The relative error of the post-Newtonian
formulas in the $r$-component of the ($\tilde{\rm S}-{\rm S}$)
force for the
radii $r_0=6M, 10M, 20M$ and $50M$. The horizontal axis is the
order of post-Newtonian expansion. The top figure shows the
convergence in the Taylor expansion and the bottom figure is the
one obtained by using the Pade approximation.} 
\label{fig:FRerr}
\end{figure}

The accuracy is actually limited by the ($\tilde{\rm S}-{\rm S}$) part.
In Fig.~\ref{fig:FRerr}, we show the convergence of the
$r$-component of the ($\tilde{\rm S}-{\rm S}$) force as a function of the
PN order for several representative orbital radii
$r_0$, so the
accuracy of the full regularized force can be read from this figure.
Here an estimator of convergence of the PN expansion is
 defined by
\begin{equation}
\Delta_{\alpha}^{\tilde{\rm S}-S}(n)
:= \left|\frac{F_\alpha^{\tilde{\rm S}-S}|_{(n)} -
F_\alpha^{\tilde{\rm S}-S}|_{(n-1)}}{F_\alpha}\right|,
\label{conv}
\end{equation}
where $F_\alpha^{\tilde{\rm S}-{\rm S}}|_{(n)}$ 
denotes the $(\tilde{\rm S}-{\rm S})$ part of
the force truncated at $n$-PN order (inclusive),
 and the denominator $F_\alpha$ denotes 
the exact (fully relativistic) self-force
including the $\tilde{\rm R}$ part. 
In practice, since it is impossible to
know the exact value of it, we use the most accurate result in our
calculation. It is found that the
convergence of PN expansion is steady even near ISCO, although it slows
down there. The convergence improves slightly by using the Pade
approximation near ISCO. Here, in the Pade approximation, we have
chosen the denominator to be quadratic in $V^2$.

Now let us compare our result with the one obtained
by Detweiler and his collaborators~\cite{Detweiler:2002gi,Diaz-Rivera:2004ik}.
They calculated the radial component of the self-force for 
various radii in units of $M=1$ and $q^2=4\pi$.
In our case, we employed a Pade approximation for
the ($\tilde{\rm S}-{\rm S}$) force accurate to 18PN order and 
used the most accurate $\tilde{\rm R}$ force in our calculation 
including the terms up to $\ell=19$.
Their results are compared with ours for $r_0=6M$ (ISCO),
 $10M$ and $20M$ in Table~\ref{tab:FR}.
As clear from it, the agreement is impressive for $r_0=10M$ and $20M$,
while there is relative error of $\sim 10^{-4}$ for $r_0=6M$.
This is consistent with the error estimate given in Fig.~\ref{fig:FRerr}.
This error may seem large, but if we use our result as a template for 
a space gravitational wave detector such as
LISA, it turns out that the error is small enough~\cite{HJNSST2}.
Thus, we conclude that our new method is capable of computing the
self-force with sufficient accuracy, and the result obtained 
to 18PN order seems accurate enough even in the limit of ISCO.

\begin{table}[ht]
\begin{center}
\tabcolsep=2mm
\doublerulesep=2mm
\renewcommand{\arraystretch}{1.4}
\begin{tabular}{c|c|c|c}
 $r_0$ & $6M$ & $10M$ & $20M$ 
\\
\hline\hline
 $F_r^{{\rm R}}(r_0)$
   &$1.676820878\times10^{-4}$ &$1.378448171\times10^{-5}\hfill$
   &$4.937905866\times10^{-7}$  
\\
\hline
Detweiler et al.&
$1.6772834\times10^{-4}\hfill$ &$1.37844828(2)\times10^{-5}$
& $4.937906\times10^{-7}\hfill$  
\end{tabular}
\caption{The $r$-component of the self-force in units of $M=1$ and
$q^2=4\pi$ obtained by our method (top column) and by Detweiler 
et al.~\cite{Detweiler:2002gi,Diaz-Rivera:2004ik} (bottom column).
 The number for $r_0=10M$ is taken from
Ref.~\cite{Detweiler:2002gi} in which they estimate the error by
a Monte Carlo simulation, and the numbers for $r_0=6M$ and $20M$
are taken from Table~I of Ref.~\cite{Diaz-Rivera:2004ik}.}
 \label{tab:FR}
\end{center}
\end{table}

\section{Conclusion}

In this paper, we have reviewed recent progress in the 
regularization and computation of the self-force acting
on a particle orbiting a black hole.

We have presented a method to regularize the self-force analytically,
employing the post Newtonian expansion.
If we were to perform regularization numerically, the fraction to be
subtracted would become closer and closer to unity as $\ell$ increases.
Apparently, this would mean a stringent requirement on numerical accuracy.
This is a clear advantage in the analytical approach over a numerical one.
Thus, as far as the Schwarzschild background
is concerned, we are almost ready to calculate
the gravitational self-force for general orbits
with a sufficiently high accuracy in post Newtonian expansion.
Research along this direction is now in progress~\cite{future}.

At the same time, however, we should mention that our analytic
method will become ineffective in the very high frequency regime.
Therefore, it seems important to develop an alternative
numerical method that may play a complimentary role.

Furthermore, although these recent developments are substantial,
they are not quite good enough because our final goal 
is to compute the regularized
self-force in the Kerr background. As mentioned in the 
introduction, there is some progress in the case of the
Kerr background~\cite{Ori:2002uv,Barack:2002mh2}.
However, we should admit that we are still at a primitive
stage in the case of the Kerr background.

Recently, extending the adiabatic orbital evolution
in a more systematic manner, Mino proposed an alternative, 
possibly much more powerful method to deal with the gravitational
reaction force to the orbit of a particle~\cite{Mino:2003yg}. 
Perhaps, we should test his idea by applying it to orbits 
in the Schwarzschild background and comparing the orbital evolution 
with the one obtained from the self-force regularization method.
 Together with computation of
the gravitational
self-force for general orbits in the Schwarzschild background,
we hope to come back to this issue in the near future.

\ack
We would like to thank L. Barack, S. Jhingan, Y. Mino, A. Ori, 
N. Sago, H. Tagoshi and T. Tanaka for fruitful collaboration on
various occasions, on which this work is based. 
We also thank all the participants of the 7th Capra meeting at the Center
for Gravitational Wave Astronomy (CGWA), University of Texas Brownsville
for useful comments and discussions.
HN is supported by JSPS Research Fellowships for Young Scientists,
No.~5919. This work was supported in part by Monbukagaku-sho
Grant-in-Aid for Scientific Research, Nos.~14047212 and 14047214,
and by the Center for Diversity and Universality in Physics, which 
is one of the 21st Century Center of Excellence programs at Kyoto 
university.


\end{document}